\documentclass[sigconf,screen,authorversion,nonacm]{acmart}

\usepackage{booktabs} 
\usepackage{amsmath,bm}
\usepackage{amsthm}
\usepackage{stmaryrd}
\usepackage{array}
\usepackage{listings}
\usepackage[shadow]{todonotes}
\usepackage{xspace}
\usepackage{paralist}

\usepackage{enumitem}
\usepackage{multirow}
\usepackage{xcolor,colortbl}
\usepackage{flushend}
\usepackage{makecell}
\usepackage{caption}
\usepackage{arydshln}
\usepackage{mdframed}
\usepackage{flushend}
\usepackage{url}
\usepackage{subcaption}
\usepackage[utf8]{inputenc}
\usepackage[T1]{fontenc}
\usepackage{microtype}
\usepackage{algorithm}
\usepackage[noend]{algpseudocode}
\usepackage[most]{tcolorbox}
\usepackage{cleveref} 


\algrenewcommand\algorithmicindent{0.5em}

\newcommand{\ignore}[1]{}

\newcommand{\nrx}[1]{}
\newcommand{\dtx}[1]{}
\newcommand{\sharonx}[1]{}
\newcommand{\revision}[1]{#1}
\definecolor{revisioncolor}{rgb}{0.69, 0.72, 0.99}
\newcommand{\revisionrow}{}

\newcommand{\ie}{i.e.,\ }
\newcommand{\eg}{e.g.,\ }

\newcommand{\etc}{etc.}

\newcommand{\klee}{KLEE\xspace}

\newcommand{\code}[1]{\texttt{#1}}

\newcolumntype{C}[1]{>{\centering\let\newline\\\arraybackslash\hspace{0pt}}m{#1}}
\newcolumntype{L}[1]{>{\raggedright\let\newline\\\arraybackslash\hspace{0pt}}m{#1}}
\newcolumntype{R}[1]{>{\raggedleft\let\newline\\\arraybackslash\hspace{0pt}}m{#1}}

\algnewcommand\algorithmicforeach{\textbf{foreach}}
\algdef{S}[FOR]{ForEach}[1]{\algorithmicforeach\ #1\ \algorithmicdo}

\usepackage{aliascnt}

\newtheorem{example}{Example}

\newtheorem{theorem}{Theorem}[section]

\theoremstyle{plain}

\newaliascnt{lemma}{theorem}
\newtheorem{lemma}[lemma]{Lemma}
\aliascntresetthe{lemma}

\newaliascnt{corollary}{theorem}

\aliascntresetthe{corollary}

\theoremstyle{definition}

\newaliascnt{definition}{theorem}
\newtheorem{definition}[definition]{Definition}
\aliascntresetthe{definition}

\newcommand{\mvfunc}{\textit{merge\_var}}

\newcommand{\qfabv}{\textit{QFABV}\xspace}
\newcommand{\ite}{\textit{ite}\xspace}
\newcommand{\select}{\textit{select}\xspace}

\newcommand{\uclibc}{\textit{uClibc}}
\newcommand{\kleeuclibc}{\textit{klee-uclibc}}
\newcommand{\libtasn}{\textit{libtasn1}\xspace}
\newcommand{\libpng}{\textit{libpng}\xspace}
\newcommand{\libosip}{\textit{libosip}\xspace}
\newcommand{\wget}{\textit{wget}\xspace}

\newcommand{\apr}{\textit{apr}\xspace}
\newcommand{\jsonc}{\textit{json-c}\xspace}
\newcommand{\busybox}{\textit{busybox}\xspace}

\newcommand{\base}{\textit{BASE}\xspace}
\newcommand{\smopt}{\textit{SMOpt}\xspace}
\newcommand{\cfg}{\textit{CFG}\xspace}

\newcommand{\patternabv}{\textit{PAT}\xspace}
\newcommand{\optstrip}{\textit{S}\xspace}
\newcommand{\optdup}{\textit{D}\xspace}
\newcommand{\optrepair}{\textit{R}\xspace}

\newcommand\x{{\small$\times$}\xspace}

\newcommand{\boundvar}{i}
\newcommand{\freevar}{i}
\newcommand{\bodyform}{\psi}

\newcommand{\qfform}{\theta}
\newcommand{\clause}{c}
\newcommand{\qclause}{\forall \boundvar.\ 1 \leq \boundvar \leq k \rightarrow \bodyform}

\newcommand{\qfclause}{\theta}

\newcommand{\arrays}{\mathrm{q\_arrays}}
\newcommand{\reads}{\mathrm{reads}}
\newcommand{\qreads}{\mathrm{q\_reads}}
\newcommand{\termsform}{\mathrm{terms}}

\newcommand{\conflicts}{\mathit{conflicts}}
\newcommand{\map}{\mathit{map}}

\newcommand{\fv}[1]{\mathrm{free\_vars(#1)}}

\newcommand{\updateselect}{\mathrm{update\_select}}

\newcommand{\quantified}[1]{\mathrm{q}(#1)}
\newcommand{\qfree}[1]{\mathrm{qf}(#1)}
\newcommand{\semanticterm}[1]{\tilde{#1}}

\newcommand{\Nat}{\mathbb{N}}

\newcommand{\treepath}{\pi}
\newcommand{\treepc}{\ensuremath{\mathit{tpc}}}
\newcommand{\treepctail}{\ensuremath{\overline{\mathit{tpc}}}}

\newcommand{\hashwhite}{\textit{W}}
\newcommand{\hashred}{\textit{R}}
\newcommand{\hashgreen}{\textit{G}}
\newcommand{\hashblue}{\textit{B}}
\newcommand{\hashyellow}{\textit{Y}}

\newcommand{\smtcomputemodel}{\textsc{smt-compute-model}}
\newcommand{\computemodel}{\textsc{compute-model}}
\newcommand{\strip}{\textsc{strip}}
\newcommand{\duplicatemodel}{\textsc{duplicate}}
\newcommand{\repairmodel}{\textsc{repair}}

\newcommand{\eqsyn}{\doteq}

\newcommand{\introfunc}{\code{memspn}\xspace}

\newcommand\aaccess[2]{#1[#2]}

\newcommand{\cstring}[1]{\texttt{#1}}

\newcommand{\textbfit}[1]{\textbf{\textit{#1}}}
\newcommand{\rqanswer}[2]{\textbfit{RQ#1 Answer:} \textit{#2}}

\newcommand{\citeapp}[1][]{Appendix #1}

\newcommand{\pbreak}{}

\begin{document}

\title{State Merging with Quantifiers in Symbolic Execution}

\author{David Trabish}
\affiliation{%
  \institution{Tel Aviv University}
  \city{Tel Aviv}
  \country{Israel}
}
\email{davivtra@post.tau.ac.il}

\author{Noam Rinetzky}
\affiliation{
  \institution{Tel Aviv University}
  \city{Tel Aviv}
  \country{Israel}
}
\email{maon@cs.tau.ac.il}

\author{Sharon Shoham}
\affiliation{
  \institution{Tel Aviv University}
  \city{Tel Aviv}
  \country{Israel}
}
\email{sharon.shoham@cs.tau.ac.il}

\author{Vaibhav Sharma}
\affiliation{
  \institution{University of Minnesota}
  \city{Minneapolis}
  \country{USA} 
}
\email{vaibhav@umn.edu}

\begin{abstract}
We address the problem of constraint encoding explosion which hinders the applicability of state merging in symbolic execution.
Specifically, our goal is to reduce the number of disjunctions and \emph{if-then-else} expressions introduced during state merging.
The main idea is to dynamically partition the symbolic states
into merging groups according to a similar uniform structure detected in their path constraints,
which allows to efficiently encode the merged path constraint and memory using quantifiers.
To address the added complexity of solving quantified constraints, we propose a specialized 
solving procedure that reduces the solving time in many cases.
Our evaluation shows that our approach can lead to significant performance gains.


	
\end{abstract}

\maketitle



\section{Introduction}\label{sec:introduction}

Symbolic execution is a powerful program analysis technique that has gained significant attention over the last years in both academic and industrial areas,
including software engineering, software testing, programming languages, program verification, and cybersecurity.
It lies at the core of many applications, such as
high-coverage test generation~\cite{klee,exe,spf},
bug finding~\cite{klee,sage},
debugging~\cite{bugredux},
automatic program repair~\cite{semfix,mechtaev2016angelix},
cross checking~\cite{klee-fp,loops:pldi19},
and side-channel analysis~\cite{pasareanu2016multi,brennan2018symbolic,brotzman2019casym}.
In symbolic execution,
the program is run with an unconstrained \textit{symbolic} input,
rather than with a concrete one.
Whenever the execution reaches a branch that depends on the symbolic input,
an SMT solver~\cite{de2011satisfiability} is used to determine the feasibility of each branch side,
and the feasible paths are further explored while updating their path constraints with the corresponding constraints.
Once the execution of a given path is completed,
the solver provides a satisfying assignment for the corresponding path constraints, from which a concrete test case that replays that path can be generated.


A key remaining challenge in symbolic execution is path explosion~\cite{CadarSen13}.
State merging~\cite{vexsym,merging:pldi12} is a well-known technique for mitigating this problem,
which trades the number of explored paths with the complexity of the generated constraints.
More specifically, merging multiple symbolic states results in a symbolic state
where the path constraint is expressed using a disjunction of constraints, and the memory contents are expressed using \ite (if-then-else) expressions.

Unfortunately, the introduction of disjunctive constraints and \ite expressions
makes constraint solving harder and slows down the exploration,
especially when the number of states being merged is high.
Consider, for example, the function \introfunc from \Cref{fig:example}
which is based on the implementation of \code{strspn} in \textit{uClibc}~\cite{uclibc}.\footnote{\code{strspn} receives null-terminated buffers, slightly complicating the presentation.}
\introfunc  receives a buffer \code{s},   the size of the buffer  \code{n}, and a string \code{chars},
and returns the size of the initial segment of \code{s} which consists entirely of characters in \code{chars}.
Suppose that \introfunc is called with a symbolic buffer \code{s}, a symbolic size \code{n} bounded by some constant $m$,
and the constant string \cstring{"a"}.
\revision{
The exploration of the loop at lines~\ref{line:loop-start}-\ref{line:loop-end} results in $O(m)$ symbolic states.
If we merge these symbolic states,
then the encoding of the merged symbolic state,
which records, among others, the path constraint and the value of variable \code{count},
is of size at least linear in $m$.
Now, suppose that the merged return value of \introfunc is used later,
for example, in the parameter \code{s} in another call of \introfunc.
In that case,
if we perform a similar merging operation,
then the encoding of the merged symbolic state will be of size at least quadratic in $m$
since the merged value propagates to the path constraints.}
Such \textit{encoding explosion} is typically encountered during the analysis of real-world programs,
thus drastically limiting the effectiveness of state merging in practice.

\begin{figure}
\caption{Motivating example.}
\label{fig:example}
\begin{lstlisting}[linewidth=.99\columnwidth,mathescape=true]
int memspn(char *s, size_t n, char *chars) {
    char *p = chars; int count = 0;
    while (*p && count < n) { /*@ \label{line:loop-start} @*/
        if (*p == s[count]) { /*@ \label{line:if-statement} @*/
            count++; p = chars;/*@ \label{line:match} @*/
        } else
            p++;
    } /*@ \label{line:loop-end} @*/
    return count;
}
\end{lstlisting}
\end{figure}

We propose a state merging approach that reduces the encoding complexity of the path constraints and the memory contents,
while preserving soundness and completeness w.r.t. standard symbolic execution.
At a high level,
our approach takes as an input the execution tree~\cite{king76}, which characterizes the symbolic branches occurring during the symbolic execution of the analyzed code fragment,
and dynamically detects regular patterns in the path constraints of the symbolic states in the tree,
which allows us to partition them into merging groups of states whose path constraints have a similar \emph{uniform} structure.
This enables us to encode the merged path constraints using quantified formulas,
which in turn may also simplify the encoding of \ite expressions representing the merged memory contents.

We observed that the generic method employed by the SMT solver to solve the resulting quantified queries
often leads to subpar performance compared to the solving of the quantifier-free variant of the queries.
To address this, we propose a specialized solving procedure that leverages the particular structure of the generated quantified queries,
and resort to the generic method only if our approach fails.

We implemented our approach on top of \klee~\cite{klee} and evaluated it on real-world benchmarks.
Our experiments show that our approach can have significant performance gains compared to state merging and standard symbolic execution.

\pbreak


\section{Preliminaries}
\label{sec:preliminaries}

\textbf{\textit{State Merging.}}
\label{sec:state-merging}
\revision{A \emph{symbolic state} $s$ consists of
\begin{inparaenum}[(i)]
\item a \emph{path constraint} $s.pc$,
\item a \emph{symbolic store} $s.\mathit{mem}$ that associates variables\footnote{
For simplicity, we do not describe the handling of stack variables and heap-allocated objects.
Our implementation supports both. 
} $V$ with symbolic expressions obtained from the symbolic inputs,
\item and an \emph{instruction counter} $s.\mathit{ic}$.
\end{inparaenum}}
Symbolic states are \textit{merge-compatible}
if they have the same instruction counter and contain the same variables in their stores.

\begin{definition}
\label{def:state-merging}
The merged symbolic state resulting from the merging of the merge-compatible symbolic states $\{s_i\}_{i = 1}^{n}$
is the symbolic state $s$ defined as follows:
\[
\begin{array}{l}
s.\mathit{ic} \triangleq s_1.ic, \ s.\mathit{pc} \triangleq \bigvee_{i = 1}^{n} s_i.pc, \\
s.\mathit{mem} \triangleq \lambda v \in V. \ \mvfunc(\{s_i\}_{i = 1}^{n}, v)
\end{array}
\]
where the merged value of a variable $v$ is defined by:
\[
\begin{array}{l}
\hspace{-2ex} \mvfunc(\{s_i\}_{i = 1}^{n}, v) \triangleq \\
\hspace{0ex} \ite(s_1.pc, s_1.\mathit{mem}(v), \\
\hspace{3ex} \ite(\ldots, \ite(s_{n - 1}.pc, s_{n - 1}.\mathit{mem}(v), s_n.\mathit{mem}(v))))
\end{array}
\]
\end{definition}

State merging is applied on a given code fragment, typically a loop or a function.
Once the symbolic exploration of the code fragment is complete,
the resulting symbolic states are partitioned into (merge-compatible) merging groups.
Then, each merging group is transformed into a single merged symbolic state.
Finally, the resulting merged symbolic states are added to the state scheduler~\cite{klee}
of the symbolic execution engine to continue the exploration.

\begin{figure}
\centering
\includegraphics[width=240pt,clip]{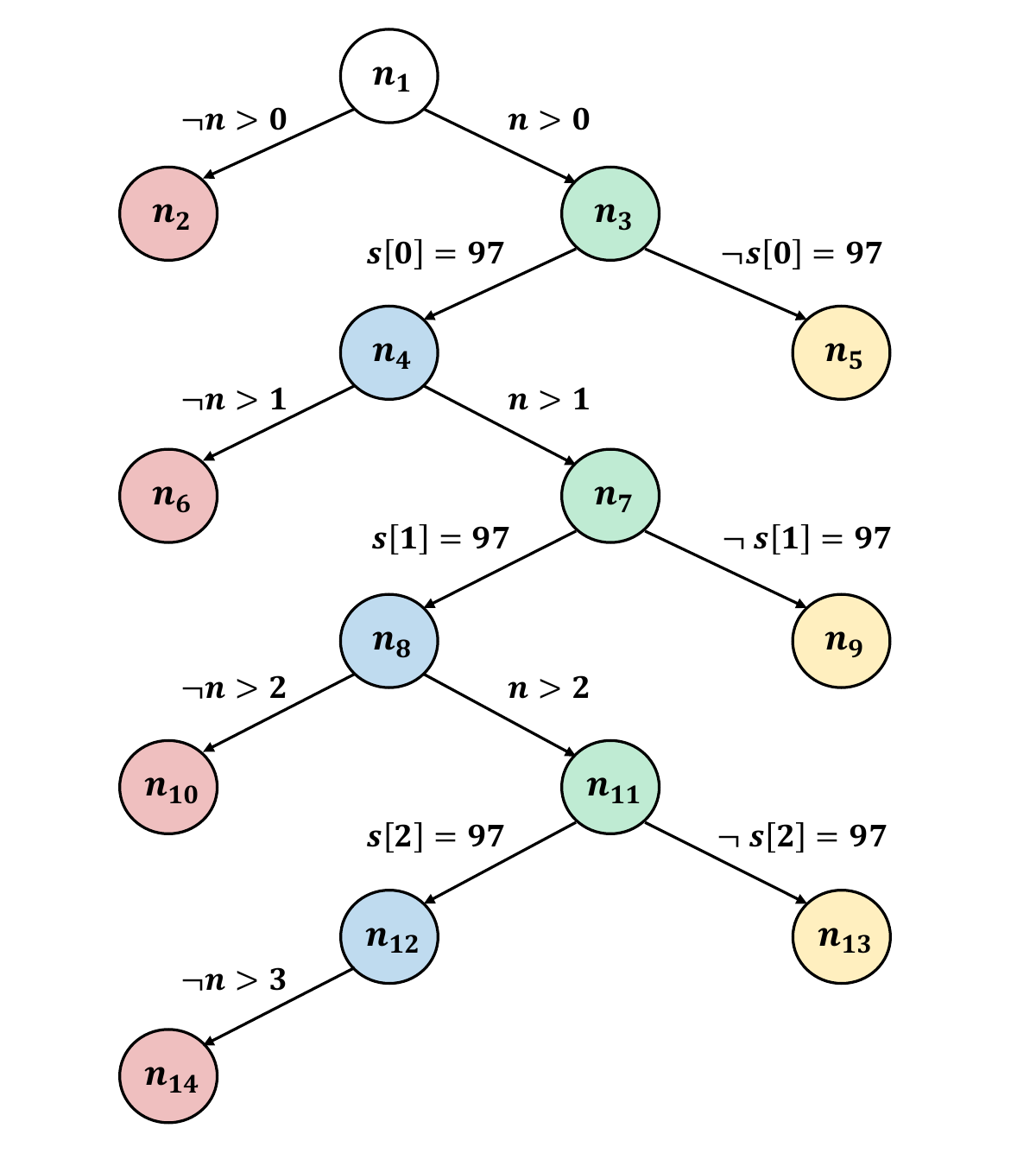}
\caption{\revision{The execution tree of the loop from \Cref{fig:example} when \code{chars} is set to \code{"a"}. (Recall that the ASCII code of \code{a} is 97.)}}
\label{fig:execution-tree}
\end{figure}

\textbf{\textit{Execution Trees.}}
An \emph{execution tree}~\cite{king76} is a tree where
every node $n$ is associated with a symbolic state $n.s$ and a symbolic condition $n.c$ corresponding to the taken branch
such that the conditions associated with any two sibling nodes are mutually inconsistent and the condition of the root node is $\mathit{true}$.
The execution tree characterizes the analysis of an arbitrary code fragment, which is not necessarily the whole program.
The root node corresponds to the symbolic state that reached the entry point of the code fragment,
and the leaf nodes correspond to the symbolic states that completed the analysis of the code fragment.
\revision{
For example,
consider the symbolic execution of \introfunc (\Cref{fig:example}) with a symbolic buffer \code{s}, a symbolic size \code{n}, and \cstring{"a"},
where \code{n} is bounded by 3. 
The corresponding execution tree is depicted in~\Cref{fig:execution-tree},
where the symbolic condition associated with each node is depicted on the incoming edge of the node.
The node $n_1$ corresponds to the initial symbolic state (\ie $n_1.s.pc \triangleq n \leq 3$),
the nodes $n_2$, $n_6$ $n_{10}$, and $n_{14}$ correspond to paths where \code{s} is comprised of only \code{a} characters,
and the nodes $n_5$, $n_9$, and $n_{13}$ correspond to paths where \code{s} contains a non-\code{a} character.
For now, ignore the color of the nodes.}


Given an execution tree $t$ with root $r$,
we denote the sequence of nodes on the path from node $n_1$ to node $n_k$ in $t$ by $\treepath_t(n_1, n_k)$
and write $\treepath_t(n_k)$ when $n_1$ is the root $r$.
Given a path $\treepath_t(n_1, n_k)=[n_1, n_2, ..., n_k]$ in $t$,
we define its \emph{tree path condition} ($\treepc$) and \emph{tree path condition tail} ($\treepctail$):
\[ \treepc_t(n_1, n_k) \triangleq n_1.c \land \treepctail_t(n_1, n_k) \qquad \treepctail_t(n_1, n_k) \triangleq \bigwedge_{1 < i \leq k} n_i.c \]
We write $\treepc_t(n) \triangleq \treepc_t(r, n) $ and $\treepctail_t(n) \triangleq \treepctail_t(r, n)$ as shorthands.
We omit the tree subscript when it is clear from the context.
For example, in the execution tree depicted in \Cref{fig:execution-tree}:
\begin{eqnarray*}
\treepath(n_3, n_7) & \triangleq & [n_3, n_4, n_7] \\
\treepc(n_3, n_7) & \triangleq & n > 0 \land s[0] = 97 \land n > 1 \\
\treepctail(n_3, n_7) & \triangleq & s[0] = 97 \land n > 1
\end{eqnarray*}

An execution tree $t$ with root $r$ is \textit{valid} if
$n.s.pc = r.s.\mathit{pc} \land \treepc(n)$ for every node $n$.
Note that $r.s$ is not necessarily the initial symbolic state of the whole program,
so $\treepc(n)$ is a suffix of the path constraints.
From now on, we assume that all trees are valid.

\textbf{\textit{Logical Notations.}}
We encode symbolic path constraints and memory contents in first-order logic modulo theories
using \emph{formulas} and \emph{terms}, respectively.
A term is either a constant, a variable, or an application of a function to terms.
A formula is either an application of a predicate symbol to terms or obtained by applying boolean connectives or quantifiers to formulas.
Let $\varphi$, $\varphi'$ be formulas and $m$ a model.
We write $\varphi \equiv \varphi'$ to note that $\varphi$ and $\varphi'$ are semantically equivalent
and $\varphi \eqsyn \varphi'$ to note that they are syntactically equal.
We write $m \models \varphi$ to note that $m$ is a model of $\varphi$.
For a term $t$, we denote by $m(t)$ the value assigned by $m$ to $t$,
and we write $t_1 \equiv t_2$ to denote that $m(t_1) = m(t_2)$ in any model $m$.
We use the standard theory of arrays~\cite{SBDL01} and write $\aaccess{a}{e}$ as a shorthand for $\select(a,e)$.

\pbreak


\section{State Merging with Quantifiers}
\label{sec:pattern-based}

\begin{table*}[t]
\caption{
A regular partitioning of the leaf nodes of the execution tree in \Cref{fig:execution-tree}, and the resulting merged states.
}
\label{table:patterns}
\centering
\renewcommand{\arraystretch}{1.3}

{\footnotesize
\begin{tabular}{|l|l|l|l|}
\cline{1-3}
\multicolumn{1}{|c|}{\textbf{Regular Pattern}} &
\multicolumn{1}{c|}{\textbf{Regular Partition}} &
\multicolumn{1}{c|}{\textbf{Pattern-Based Merged States}} \\
\hline

$(\hashwhite, \hashgreen\hashblue, \hashgreen\hashyellow)$ &
$\{n_5, n_9, n_{13}\}$ &
$\textit{formula pattern}: (\mathit{true}, n > x - 1 \land s[x - 1] = 97, n > x \land \neg s[x] = 97)$ \\
\cline{3-3}
& &
$\mathit{pc}: n \leq 3 \land 0 \leq k \leq 2 \land (\forall i. 1 \leq i \leq k \rightarrow n > i - 1 \land s[i - 1] = 97) \land (n > k \land \neg s[k] = 97)$ \\
& &
$\mathit{mem}: [\,
\code{count} \mapsto k, \quad
\code{p} \mapsto \mathit{chars} + 1, \quad
\code{s} \mapsto \mathit{s}, \quad
\code{n} \mapsto \mathit{n}, \quad
\code{chars} \mapsto \mathit{chars}
\,] $\\
\hline

$(\hashwhite, \hashgreen\hashblue, \hashred)$ &
$\{n_2, n_6, n_{10}, n_{14}\}$ &
$\textit{formula pattern}: (\mathit{true}, n > x - 1 \land s[x - 1] = 97, n \leq x)$ \\
\cline{3-3}
& &
$\mathit{pc}: n \leq 3 \land 0 \leq k \leq 3 \land (\forall i. 1 \leq i \leq k \rightarrow n > i - 1 \land s[i - 1] = 97) \land \neg n > k $ \\
& &
$\mathit{mem}: [\,
\code{count} \mapsto k, \quad
\code{p} \mapsto \mathit{chars}, \quad
\code{s} \mapsto \mathit{s}, \quad
\code{n} \mapsto \mathit{n}, \quad
\code{chars} \mapsto \mathit{chars}
\,] $\\
\hline

\end{tabular}
}
\end{table*}

In this section, we describe our approach for state merging with quantifiers.
We start with a motivating example and subsequently formalize our approach.

\textbf{\textit{Motivating Example.}}
Consider the symbolic states associated with the nodes $n_5$, $n_9$, and $n_{13}$ from the execution tree in~\Cref{fig:execution-tree},
whose tree path conditions, \ie, $\treepc(n_5)$, $\treepc(n_9)$, and $\treepc(n_{13})$, are:
\[
\begin{array}{l}
\hspace{-1.3ex} n > 0 \land \neg s[0] = 97 \\
\hspace{-1.3ex} n > 0 \land s[0] = 97 \land n > 1 \land \neg s[1] = 97 \\
\hspace{-1.3ex} n > 0 \land s[0] = 97 \land n > 1 \land s[1] = 97 \land n > 2 \land \neg s[2] = 97 \\
\end{array}
\] 
The path constraint of the initial symbolic state ($n_1.s$) is $n \leq 3$,
so applying standard state merging (\Cref{def:state-merging}) on the symbolic states of the nodes above
will result in a symbolic state whose path constraint is equivalent to:
\[ n \leq 3 \land (\treepc(n_5) \lor\treepc(n_9) \lor \treepc(n_{13})) \]
Note, however, that each of the disjuncts above has the following uniform structure:
It uses $k$ formulas (for $k = 0, 1, 2$) of the form $n > \_ \land s[\_] = 97$
to encode that the size of the buffer ($n$) is big enough to contain $k$ consecutive occurrences of \code{a} characters,
and another formula $n > k \land \neg s[k] = 97$.
This uniformity is exposed when rewriting each disjunct using universal quantifiers as follows:
\[
\begin{array}{l}
\big(\forall i. 1 \leq i \leq 0 \rightarrow  n > i - 1 \land s[i - 1] = 97 \big) \land n > 0 \land \neg s[0] = 97 \\
\big(\forall i. 1 \leq i \leq 1 \rightarrow  n > i - 1 \land s[i - 1] = 97 \big) \land n > 1 \land \neg s[1] = 97 \\
\big(\forall i. 1 \leq i \leq 2 \rightarrow  n > i - 1 \land s[i - 1] = 97 \big) \land n > 2 \land \neg s[2] = 97 \\
\end{array}
\]
To exploit the common structure of the rewritten disjuncts,
we can introduce an auxiliary variable ($k$) and obtain an \emph{equisatisfiable} merged path constraint\footnote{
Note that $(k = 0 \lor k = 1 \lor k = 2)$ can be rewritten as $0 \leq k \leq 2$.
}:
\[
\begin{array}{l}
n \leq 3 \land (k = 0 \lor k = 1 \lor k = 2) \land {} \\
\big(\forall i. 1 \leq i \leq k \rightarrow n > i - 1 \land s[i - 1] = 97 \big) \land {} \\
(n > k \land \neg s[k] = 97)
\end{array}
\]

The auxiliary variable allows us to achieve similar savings in the encoding of the merged memory contents.
Consider, for example,  the variable \code{count}.
Its value in the symbolic states corresponding to $n_5$, $n_9$, and $n_{13}$ is $0$, $1$, and $2$, respectively,
so its merged value with standard state merging is:
\[ \ite(\treepc(n_5), 0, \ite(\treepc(n_9), 1, 2)) \]
Note, however, that with the rewritten merged path constraint,
the path constraints of the symbolic states corresponding to $n_5$, $n_9$, and $n_{13}$ are now correlated with the values of $k$: $0$, $1$, and $2$.
As the values of \code{count} can be encoded as a function of those values,
we can simply rewrite the complex \ite expression above to  $k$.


\textbf{\textit{Our Approach.}}
Our goal is to reduce the number of disjunctions and \ite expressions introduced in standard state merging.
Given a set of merge-compatible symbolic states, our state merging approach works as follows.
First, we compute partitions of symbolic states based on the similarity of the path constraints (\Cref{sec:regular-pattern}).
Then, for each partition,
we attempt to synthesize the merged symbolic state using universal quantifiers (\Cref{sec:pattern-based-state-merging,sec:synthesis}),
and resort to standard state merging if that fails.

\subsection{Partitioning Merging Groups via Regular Patterns}
\label{sec:regular-pattern}

To identify similarity between symbolic states, we use the execution tree of the analyzed code fragment.
Recall that the symbolic states in each merging group are associated with leaf nodes and respective paths in the execution tree.
We abstract each path to a sequence of numbers using a specialized \emph{hash function},
which allows us to detect similarity between paths based on a shared regular pattern.
\begin{definition}
\label{def:hash}
A \emph{hash function} $h$ maps constraints (formulas) to numbers ($\Nat$).
We say that $h$ is \emph{valid} for an execution tree $t$ if for any two sibling nodes $n_1$ and $n_2$:
\[ h(n_1.c) \neq h(n_2.c) \]
\end{definition}

In the sequel,
we assume a fixed arbitrary valid execution tree $t$ and a fixed arbitrary valid hash function $h$ for $t$.\footnote{
In practice, we use a hash function that distinguishes between a condition and its negation,
effectively ensuring validity for any execution tree.
}
We now extend   $h$ to paths as follows:
\begin{definition}
The hash of a path $\treepath(n_1, n_k) = [n_1, ..., n_k]$ in $t$ is defined as follows:
\[ h(\treepath(n_1, n_k)) \triangleq h(n_1.c) h(n_2.c) \ldots h(n_k.c) \in \mathbb{N}^{*} \]
\end{definition}

Note that the validity of $h$ ensures that every path in $t$ is identified uniquely by its hash value.


\begin{definition}
A \textit{regular pattern} is a tuple $(\omega_1, \omega_2, \omega_3)$,
where $\omega_1, \omega_2, \omega_3 \in \Nat^*$ are words (sequences) of numbers.
Given leaf nodes $\{n_j\}_{j = 1}^{n}$ in $t$, and numbers $\{k_j\}_{j = 1}^{n} \subseteq \Nat$,
we say that $\{(n_j, k_j)\}_{j = 1}^{n}$ \emph{match} the regular pattern $(\omega_1, \omega_2, \omega_3)$ if for every $j = 1, ..., n$:
$h(\treepath(n_j)) = \omega_1\omega_2^{k_j}\omega_3$\,.
\end{definition}



\begin{definition}
A set of leaf nodes $\{n_j\}_{j = 1}^{n}$ in $t$ is called a \emph{regular partition}
if there exists a regular pattern $(\omega_1, \omega_2, \omega_3)$ and a set $\{k_j \}_{j = 1}^{n} \subseteq \Nat$
such that $\{(n_j, k_j)\}_{j = 1}^{n}$ match that pattern.
A \emph{regular partitioning} of leaf nodes in $t$ is a partitioning into disjoint regular partitions.
\end{definition}

\begin{example}
Consider a hash function $h$ that operates on the abstract syntax tree (AST) of a formula
and assigns the same pre-defined value to all the constant numerical terms.
Such a hash function ensures that formulas with a similar shape will be assigned the same hash value, for example:
\[ 
\begin{array}{c}
h(n > 0) = h(n > 1) = h(n > 2) \\
h(s[0] = 97 ) = h(s[1] = 97)
\end{array}
\] 
\Cref{fig:execution-tree} shows the resulting hash values of the nodes in the execution tree.
For simplicity, we visualize every hash value as a distinct color:
white (\hashwhite), red (\hashred), blue (\hashblue), green (\hashgreen), and yellow (\hashyellow).
Here, 
$\{(n_5, 0), (n_9, 1), (n_{13}, 2)\}$ match the regular pattern $(\hashwhite, \hashgreen\hashblue, \hashgreen\hashyellow)$ since:
\[
h(\treepath(n_5)) = \hashwhite\hashgreen\hashyellow, \
h(\treepath(n_9)) = \hashwhite\hashgreen\hashblue\hashgreen\hashyellow, \
h(\treepath(n_{13})) = \hashwhite\hashgreen\hashblue\hashgreen\hashblue\hashgreen\hashyellow
\]

A (possible) regular partitioning of the leaf nodes in~\Cref{fig:execution-tree} is given in~\Cref{table:patterns},
which shows in the two leftmost columns the regular patterns and their corresponding regular partitions.
\end{example}

In the following sections,
we show how given a regular partition and its corresponding regular pattern,
we can synthesize the resulting merged symbolic state using quantifiers.

\subsection{Pattern-Based State Merging}
\label{sec:pattern-based-state-merging}

A regular pattern indicates the potential existence of a uniform structure in the path conditions of the symbolic states in the associated regular partition.
We formalize this intuition using \emph{formula patterns}.
\begin{definition}
A \textit{formula pattern} is a tuple $(\varphi_1, \varphi_2(x), \varphi_3(x))$,
where $\varphi_1$ is a closed formula, and $\varphi_2(x)$ and $\varphi_3(x)$ are formulas with a free variable $x$.
We say that $\{(n_j, k_j)\}_{j = 1}^{n}$ \emph{match} the formula pattern $(\varphi_1, \varphi_2(x), \varphi_3(x))$,
if for every $j = 1, ..., n$:
\[ \treepc(n_j) \eqsyn \varphi_1 \land \big( \bigwedge_{i = 1}^{k_j}\varphi_2[i/x] \big) \land \varphi_3[k_j/x] \]
\end{definition}

The uniform structure exposed by formula patterns enables us to perform state merging with quantifiers:
\begin{definition}
\label{def:pattern-based-state-merging}
Let $\{n_j\}_{j = 1}^{n}$ be a set of leaf nodes in $t$
such that $\{n_j.s\}_{j = 1}^{n}$ are merge-compatible and
$\{(n_j, k_j)\}_{j = 1}^{n}$ match the formula pattern $(\varphi_1, \varphi_2(x), \varphi_3(x))$.
The \emph{pattern-based merged symbolic state} of $\{n_j.s\}_{j = 1}^{n}$ is a symbolic state $s$ whose path constraint, $s.pc$, is:
\[ r.s.pc \land (\bigvee_{j = 1}^{n} k = k_j) \land \varphi_1 \land (\forall i. \ 1 \leq i \leq k \rightarrow \varphi_2[i/x]) \land \varphi_3[k/x] \]
where $k$ is a fresh constant, $i$ is a fresh variable, and $r$ is the root of~$t$.

The symbolic store of $s$ is defined as follows. 
For every variable $v$,
if there exists a term $t(x)$ with a free variable $x$ such that $t[k_j/x] \eqsyn n_j.s.mem(v)$ for every $j = 1,\dots,n$,
then the value of $v$ is encoded as $s.\mathit{mem}(v) \triangleq t[k/x]$.
Otherwise, $s.\mathit{mem}(v) \triangleq \mvfunc(\{n_j.s\}_{j = 1}^{n}, v)$ (\Cref{def:state-merging}).
\end{definition}

Pattern-based state merging is sound and complete w.r.t.\ standard state merging.
This is formalized in the following theorem:
\begin{theorem}
\label{theorem:state-merging}
Under the premises of \Cref{def:pattern-based-state-merging},
let $s$ be the pattern-based merged symbolic state of $\{n_j.s\}_{j = 1}^{n}$,
and let $s'$ be their merged symbolic state obtained with standard state merging (\Cref{def:state-merging}).
The following holds for any model $m$:
\begin{itemize}
\item $m \models s'\hspace{-3pt}.pc$ iff $m[k \mapsto \tilde{k}] \models s.pc$ for some $\tilde{k} \in \Nat$.
\item If $m \models s.pc$ then $m(s'\hspace{-3pt}.\mathit{mem}(v)) = m(s.\mathit{mem}(v))$ for every variable $v$.
\end{itemize}
\end{theorem}

\begin{example}
Consider the regular partition $\{n_5, n_9, n_{13}\}$ shown in the first row of \Cref{table:patterns}.
The formula pattern $(true, n > x - 1 \land s[x - 1] = 97, n > x \land \neg s[x] = 97)$
is matched by $(\{(n_5, 0), (n_9, 1), (n_{13}, 2)\}$.
The merged symbolic state induced by that formula pattern is shown in the rightmost column in~\Cref{table:patterns} (\textit{pc} and \textit{mem}).
Note that for the variable \code{count}, the term $t(x) \triangleq x$ satisfies:
\[ t[0/x] = 0, t[1/x] = 1, t[2/x] = 2 \]
so the merged value of that variable can be simplified to $k$.
The merging of the other variables is rather trivial as the symbolic states being merged agree on their values.

\end{example}

\subsection{Synthesizing Formula Patterns}
\label{sec:synthesis}

So far, we have yet to discuss how formula patterns are obtained.
We now describe an approach that attempts to synthesize a formula pattern given a regular pattern and its associated regular partition.
As explained in~\Cref{sec:pattern-based-state-merging},
this enables us to perform state merging with quantifiers.

\sharonx{I doubt that someone remembers this at this point. Consider moving it to here (if that's the first place where it's important) or remind the reader; advantage of moving it to here is that it will be clearer why we need it.}

Our hash function $h$, which we assume to be valid for $t$ (\Cref{def:hash}), has the following useful property:
\begin{lemma}
\label{path-word-correspondence}
The following holds for any two nodes $n_1$ and $n_2$ in $t$:
\begin{enumerate}
\item If $h(\treepath(n_1)) = h(\treepath(n_2))$ then $n_1 = n_2$.
\item If $h(\treepath(n_1))$ is a prefix of $h(\treepath(n_2))$, then there is a single path $\treepath(n_1, n_2)$ in $t$.
\end{enumerate}
\end{lemma}

Accordingly, we define:

\begin{definition}
Let $\omega_1, \omega_2 \in \Nat^{*}$ be two words such that:
\[ h(\treepath(n_1)) = \omega_1, \ \ \  h(\treepath(n_2)) = \omega_1\omega_2 \]
for some nodes $n_1$ and $n_2$ in $t$.
Then we define:
\[ extract(\omega_1) \triangleq \treepctail(n_1), \ extract(\omega_1, \omega_1\omega_2) \triangleq \treepctail(n_1, n_2) \]
which gives us the tree path condition tails associated with the paths $\treepath(n_1)$ and $\treepath(n_1,n_2)$, respectively. 
(Note that \Cref{path-word-correspondence} ensures 
that $n_1$ and $n_2$ are uniquely determined by $\omega_1$ and $\omega_2$.) 

\end{definition}

We use \emph{extract} to define the sufficient requirements to obtain a formula pattern from a given regular pattern.
\begin{lemma}
\label{lemma:synthesis}
Suppose that $\{(n_j, k_j)\}_{j = 1}^{n}$ match the regular pattern $(\omega_1, \omega_2, \omega_3)$.
Let $(\varphi_1,\varphi_2(x),\varphi_3(x))$ be a formula pattern that satisfies:
\[
\begin{array}{l}
\hspace{-1.3ex} \varphi_1 \eqsyn extract(\omega_1) \\
\hspace{-1.3ex} \varphi_2[i/x] \eqsyn extract(\omega_1\omega_2^{i - 1}, \omega_1\omega_2^{i}) \ \ \ (i = 1, ..., \mathit{max}\{k_j\}_{j = 1}^{n}) \\
\hspace{-1.3ex} \varphi_3[k_j/x] \eqsyn extract(\omega_1\omega_2^{k_j}, \omega_1\omega_2^{k_j}\omega_3) \ \ \ (j = 1, ..., n)
\end{array}
\]
Then $\{(n_j, k_j)\}_{j = 1}^{n}$ match $(\varphi_1,\varphi_2(x),\varphi_3(x))$.
\end{lemma}

Based on \Cref{lemma:synthesis}, we reduce the problem of finding a formula pattern to two synthesis tasks, for $\varphi_2$ and $\varphi_3$.
(Note that $\varphi_1$ is trivially obtained from the first requirement of the lemma.)
Each synthesis task has the form:
\[
\varphi[d_\ell/x] \eqsyn \psi_\ell \ \ \ (\ell = 1, ..., p)
\]
where
\begin{inparaenum}[(i)]
\item $\varphi(x)$  is the formula to be synthesized (i.e., $\varphi_2$ or $\varphi_3$),
\item $p$ is the number of equations (which is either $max\{k_j\}_{j = 1}^{n}$ in the case of $\varphi_2$ or $n$ in the case of $\varphi_3$),
\item $\{\psi_\ell\}_{\ell = 1}^{p}$ are formulas (obtained from the extracted path constraints), and
\item $\{d_\ell\}_{\ell = 1}^{p}$ are constant numerical terms (which are the $i$'s in the case of $\varphi_2$ or the $k_j$'s in the case of $\varphi_3$).

\end{inparaenum}

As synthesis is a hard problem in general,
we focus on the case where all formulas in $\{\psi_\ell\}_{\ell = 1}^{p}$ are syntactically identical up to a constant numerical term,
\ie there exists a formula $\theta(y)$ such that $ \theta[\gamma_\ell/y] \eqsyn \psi_\ell$ for some numerical constants $\{\gamma_\ell\}_{\ell = 1}^{p}$.
\sharonx{do we want to generalize it to multiple arguments?  (allowing to factor out multiple numerical terms? }
To obtain $\varphi(x)$ from $\theta(y)$, it remains to synthesize a term that will express each $\gamma_\ell$ using the corresponding $d_\ell$.
Technically, if there exists a term $t(x)$ such that: 
\[ t[d_\ell/x] \equiv \gamma_\ell \ \ \ (\ell = 1, ..., p) \]
then the desired formula $\varphi(x)$ will be given by $\theta[t(x)/y]$.
When looking for such $t(x)$, we restrict our attention to terms of the form $a \cdot x + b$
where $a$ and $b$ are constant numerical terms that must satisfy:
\[ \bigwedge_{\ell = 1}^{p} (a \cdot d_\ell + b = \gamma_\ell) \]
The existence of such $a$ and $b$ can be checked using an SMT solver.

\begin{example}
Consider again the regular pattern $(\hashwhite, \hashgreen\hashblue, \hashgreen\hashyellow)$ which is matched by $\{(n_5 , 0), (n_9 , 1), (n_{13} , 2)\}$.
We look for a formula pattern $(\varphi_1, \varphi_2(x), \varphi_3(x))$ that satisfies:
\begin{align*}
\varphi_1 &\eqsyn true \tag*{$extract(\hashwhite)$} \\
\varphi_2[1/x] &\eqsyn n > 0 \land s[0] = 97 \tag*{$extract(\hashwhite, \hashwhite\hashgreen\hashblue)$} \\
\varphi_2[2/x] &\eqsyn n > 1 \land s[1] = 97 \tag*{$extract(\hashwhite\hashgreen\hashblue, \hashwhite\hashgreen\hashblue\hashgreen\hashblue)$} \\
\varphi_3[0/x] &\eqsyn n > 0 \land \neg s[0] = 97 \tag*{$extract(\hashwhite, \hashwhite\hashgreen\hashyellow)$} \\
\varphi_3[1/x] &\eqsyn n > 1 \land \neg s[1] = 97 \tag*{$extract(\hashwhite\hashgreen\hashblue, \hashwhite\hashgreen\hashblue\hashgreen\hashyellow)$} \\
\varphi_3[2/x] &\eqsyn n > 2 \land \neg s[2] = 97 \tag*{$extract(\hashwhite\hashgreen\hashblue\hashgreen\hashblue, \hashwhite\hashgreen\hashblue\hashgreen\hashblue\hashgreen\hashyellow)$} \\
\end{align*}
Consider, for example, the formulas associated with $\varphi_2$.
First, note that they are identical up to a constant numerical term,
\eg for $\theta(y) \triangleq n > y \land s[y] = 97$:
\[ \theta[0/y] \eqsyn n > 0 \land s[0] = 97 \qquad \theta[1/y] \eqsyn n > 1 \land s[1] = 97 \]
Now we look for constant numerical terms $a$ and $b$ such that:
\[ (0 = (a \cdot x + b)[1/x]) \land (1 = (a \cdot x + b)[2/x]) \]
which is satisfied by $a \triangleq 1$ and $b \triangleq -1$, therefore:
\[ \varphi_2(x) \triangleq \theta[(x - 1)/y] \eqsyn n > x - 1 \land s[x - 1] = 97 \]
We similarly synthesize $\varphi_3(x) \triangleq n > x \land \neg s[x] = 97$.
\end{example}

If we succeeded to synthesize a formula pattern $(\varphi_1, \varphi_2(x), \varphi_3(x))$ matched by $\{(n_j, k_j)\}_{j = 1}^{n}$,
we attempt to synthesize the merged value of a variable $v$ by synthesizing a term $t(x)$ that satisfies:
\[ t[k_j/x] \eqsyn n_j.s.mem(v) \ \ (j = 1, ..., n) \]
Such terms are synthesized similarly to formula patterns.


For each regular partition shown in~\Cref{table:patterns},
we automatically synthesize the formula pattern and the induced merged symbolic state using the technique above.

The proofs for \Cref{theorem:state-merging} and the other lemmas are given in~\citeapp[A].



\section{Incremental State Merging}
\label{sec:incremental}

\revision{When symbolically analyzing code fragments that contain disjunctive conditions,
the number of generated states, as well as the size of the generated execution trees, might be exponential.
In such cases, 
the exploration of the code fragment might not terminate within the allocated time budget and
the analysis might not even reach the point where state merging, 
and pattern-based state merging in particular, 
can be applied.}

To address this issue,
we propose an \textit{incremental} approach for  state merging, 
in which we merge 
leaves in the execution tree not only with other leaves but also with internal nodes during the construction of the tree.
This allows to compress the tree as it is constructed.
Once the construction of the tree is complete, we can apply our pattern-based state merging approach on the leaves.
Technically, 
in addition to the \emph{active} symbolic states, \ie those that are stored in the current leaf nodes, 
we keep also the \emph{non-active} symbolic states, \ie those that are stored in the internal nodes.
When a new leaf $n_1$ is added to the execution tree, we search for the highest node $n_2$, \ie closest to the root,
such that $n_1.s$ and $n_2.s$ are merge-compatible and have the same symbolic store \emph{w.r.t. live variables}~\cite{dragon-book}.
We additionally require that $n_1$ is unreachable from $n_2$ to avoid infinite sequences of merges.
If such a node $n_2$ is found,
we replace $n_1$ and $n_2$ (and their subtrees) with a single merged node $n_{\mathit{new}}$ that is added
as a child of their lowest common ancestor, $n_{\mathit{lca}}$.
We fix $n_{\mathit{new}}.c \triangleq \treepctail(n_\mathit{lca}, n_1) \lor \treepctail(n_\mathit{{lca}}, n_2)$ and $n_{\mathit{new}}.s$ is the merged state of $n_1.s$ and $n_2.s$.
After the above, if a node $p$ remains with a single child $n$, 
we remove $p$, redirect its incoming edge to $n$,
and update the condition of $n$ to $n.c \land p.c$.
As we merge internal nodes,
our approach does not rely on the search heuristic to synchronize between the active symbolic states to produce successful merges.
To avoid nodes with more than two children, we require that $n_{\mathit{lca}}$ is the parent of  $n_1$ or $n_2$. (This restriction can be easily lifted.)

\begin{figure}
\centering
\includegraphics[width=240pt,clip]{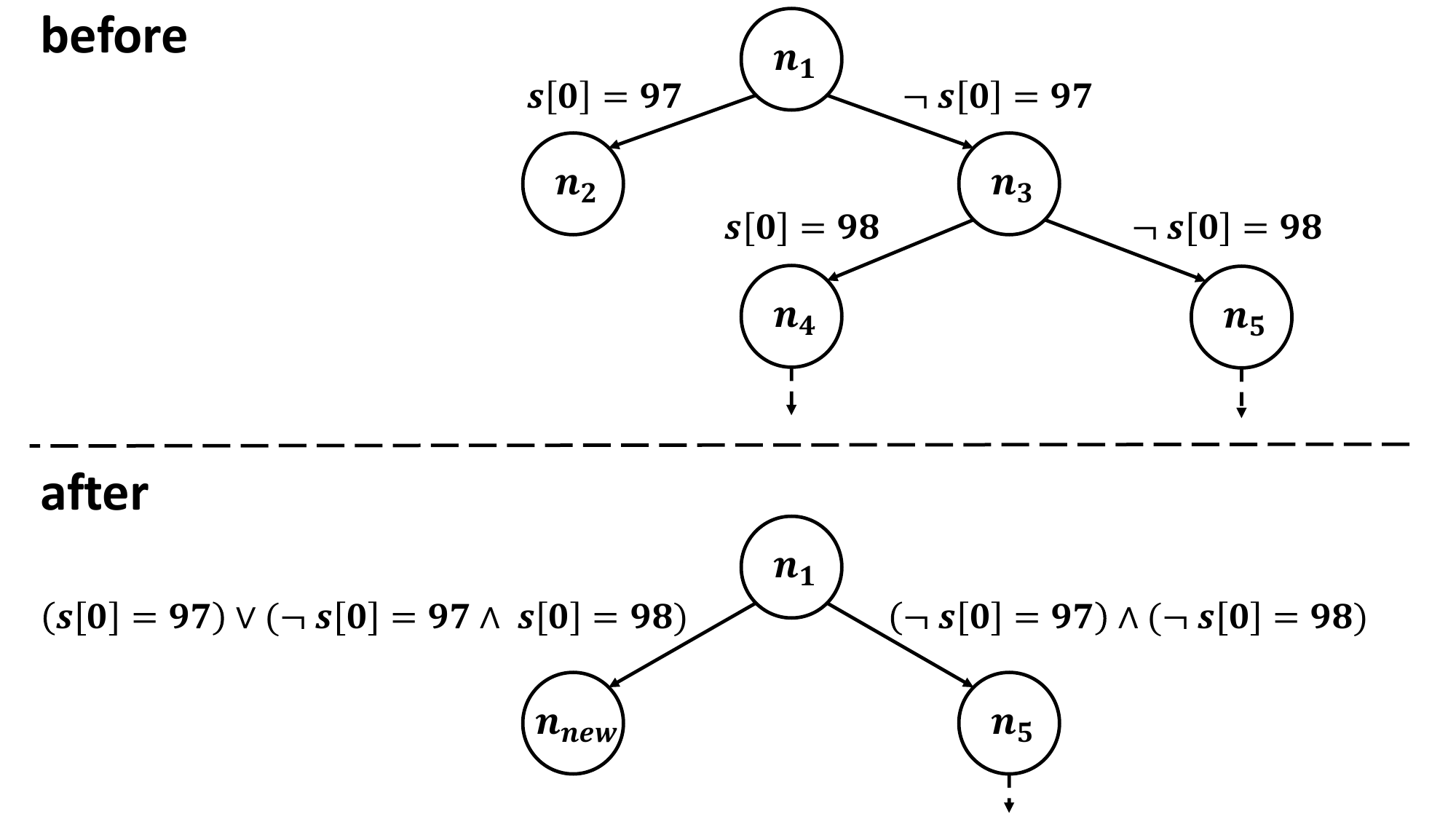}
\caption{\revision{Execution tree transformation when \code{memspn} is called with \code{chars} set to \code{"ab"}.}}
\label{fig:transformation}
\end{figure}

\begin{example}
Consider again the function \introfunc from \Cref{fig:example}.
When symbolically analyzing \introfunc while setting the value of the \code{chars} parameter to \cstring{"ab"}, instead of \cstring{"a"},
this results in an exponential execution tree.
The upper part of \Cref{fig:transformation} shows the partial execution tree
with some of the nodes that were added during the execution of the first iterations of the loop at line~\ref{line:loop-start}.
Assuming that $n_2$ is added last, we merge it with $n_4$
as the symbolic states associated with $n_2$ and $n_4$ are both located at line~\ref{line:match}
and their symbolic stores w.r.t. live variables are identical, since \code{p} is dead at this location.
We remove $n_2$ and $n_4$ together with its subtree,
and add a new node $n_{new}$ as a child of $n_1$, the lowest common ancestor of $n_2$ and $n_4$.
Then, $n_3$ is left with its own child, $n_5$, so we remove $n_3$ and appropriately update the condition of $n_5$.
This results in the execution tree shown in the lower part of \Cref{fig:transformation}.
After applying similar steps in the subsequent iterations of the loop,
the final execution tree is similar to the one from \Cref{fig:execution-tree},
and can be obtained from it by replacing
$s[i] = 97$ and $\neg s[i] = 97$ with
$s[i] = 97 \lor (\neg s[i] = 97 \land s[i] = 98)$ and $\neg s[i] = 97 \land \neg s[i] = 98$, respectively (for $i = 0,1,2$).
Now, pattern-based state merging can be applied similarly to the example given in \Cref{sec:pattern-based}.
\end{example}

\revision{The incremental state merging approach uses a standard liveness analysis~\cite{dragon-book}
to find symbolic states to be merged.
If the computed liveness results are imprecise,
our approach will not be able to find matching symbolic states
and therefore will not be able to compress the execution tree.
In that case,
our approach will only impose the overhead of maintaining snapshots of non-active symbolic states.}

\pbreak

\section{Solving Quantified Queries}\label{sec:solving}

In general, the quantified queries generated by our approach (\Cref{sec:pattern-based}) can be solved using an SMT solver that supports quantified formulas,
e.g., Z3~\cite{Z3tacas08}.
In practice, however, we observed that the generic method employed by Z3\footnote{CVC5~\cite{cvc5} and Yices~\cite{yices2} failed to solve most of our queries.} to solve such queries often leads to subpar performance
compared to the solving of the quantifier-free variant of the queries.
Hence, we devise a solving procedure that leverages the particular structure of the generated quantified formulas,
and resort to the generic method if our approach fails.

Our solving procedure assumes a closed formula $\varphi = \bigwedge \clause$
where each clause $\clause$ is either a quantifier-free formula or
a universal formula of the form $\qclause$ where $\bodyform$ is a quantifier-free formula with a free variable~$\freevar$.
Our solving procedure works in four stages:\footnote{
For the interested reader, a complete pseudo code of the solving procedure is given  in~\citeapp[B].
}

\textbf{\textit{(1) Quantifier stripping.}}
We weaken $\varphi$ into a quantifier-free formula $\varphi_{\mathit{QF}}$ by replacing quantified clauses with implied quantifier-free clauses.
Technically, each quantified clause $\qclause$ in $\varphi$ is replaced with the conjunction of the following two quantifier-free formulas\footnote{
\revision{We write $c \in \varphi$ to note that $c$ is one of the clauses of $\varphi$.}
}:
\[
(1) \ k \geq 1 \rightarrow \bodyform[1/i] \ \ \
(2) \ \bigwedge \{ \neg (1 \leq t \leq k) \mid (\neg \bodyform[t/\freevar]) \in \varphi \}
\]
Intuitively, the former provides a quantifier-free clause which partially preserves the properties imposed by the quantified clause, 
and the latter reduces the chances that the SMT solver computes a model of $\varphi_{\mathit{QF}}$ that does not satisfy $\varphi$: if $1\leq t \leq k$ then $\qclause$ demands that  $\bodyform[t/\boundvar]$ holds in any model of $\varphi$.
If the SMT solver fails to find a model for $\varphi_{\mathit{QF}}$, then $\varphi$ is also unsatisfiable.
If a model was found, we check whether it is also a model of $\varphi$.

\begin{example}\label{ex:strip}
Consider the following query, a simplification of a representative query from our experiments:
\begin{equation*}
\begin{split}
\varphi \triangleq & (s[n] = 0) \land (1 \leq k \leq 10) \land (s[k - 1] = 8) \land \\
                   & (\forall i. \ 1 \leq i \leq k \rightarrow s[i - 1] \neq 0)
\end{split}
\end{equation*}
Note that (a) the instantiation of the quantified formula using $i = 1$ results in $k \geq 1 \rightarrow s[0] \neq 0$,
and (b) $s[n] = 0$ is obtained by substituting $\neg(s[i - 1] \neq 0)[n + 1/i]$.
Thus, the weakened query obtained by quantifier stripping is given by:
\begin{equation*}
\begin{split}
\varphi_{\mathit{QF}} \triangleq & (s[n] = 0) \land (1 \leq k \leq 10) \land (s[k - 1] = 8) \land \\
                                 & (k \geq 1 \rightarrow s[0] \neq 0) \land \neg(1 \leq n + 1 \leq k)
\end{split}
\end{equation*}
The following model, for example, is a model of $\varphi_{\mathit{QF}}$:
\[ m \triangleq \{n \mapsto 7, k \mapsto 7, s \mapsto [1,0,0,0,0,0,8,0]\} \]
but, unfortunately, it is not a model of $\varphi$.
\end{example}

\textbf{\textit{(2) Assignment Duplication.}}
If $m$ is not a model of $\varphi$, we modify $m$ into a model $m_d$
which assigns to every array cell accessed by a quantified clause a value $v$ of a cell in that array that was explicitly constrained by $\varphi_{\mathit{QF}}$.
Technically, for every array $a$ accessed with an offset that depends on the quantified variable $\boundvar$ we do the following:
(1) pick an accessed offset $o$ of $a$ in $\bodyform$ such that $o$ depends on $\boundvar$,
(2) evaluate the value of $(a[o])[1/i]$ in $m$, namely $v$, and
(3) compute the concrete offsets obtained by evaluating $o[j/i]$ in $m$ (for $2 \leq j \leq m(k)$)
and modify $m$ such that the values of $a$ at these offsets are set to $v$.
Recall that the accessed cells of $a$ in $\bodyform[1/i]$ were explicitly constrained in $\varphi_{\mathit{QF}}$,
so $v$ is a good candidate to fill in all the other cells of $a$ constrained in $\varphi$.
However, this duplication is rather naive and might result in a model that does not even satisfy $\varphi_{\mathit{QF}}$.
\begin{example}\label{ex:duplicate}
Continuing \Cref{ex:strip},
we pick from the quantified clause the accessed offset $i - 1$ of the array $s$,
and update the value of $s[j]$ to $m(s[i - 1][1/i])$ for each $1 \leq j \leq 6$.
This results in the following model:
\[ m_d \triangleq \{ n \mapsto 7, k \mapsto 7, s \mapsto [1,1,1,1,1,1,1,0] \} \]
\revision{The model $m_d$ helps to satisfy the quantified clause,
but does not satisfy $\varphi$ (specifically, the clause $s[k - 1] = 8$ is violated).}
\end{example}

\textbf{\textit{(3) Model Repair.}}
If $m_d$ is not a model of $\varphi$, we further modify $m_d$ into another model, $m_r$,
which, much like $m_d$, attempts to satisfy the constraints on the contents of arrays that are imposed by $\varphi$ but omitted in $\varphi_{\mathit{QF}}$.
For every quantified clause $\qclause$, we collect all the accesses $a[o]$ where $o$ depends on $\boundvar$.
For each such access and for each $2 \leq j \leq m(k)$,
we compute the concrete offset  obtained by evaluating $o[j/i]$ in $m_d$ 
and strengthen $\varphi_{\mathit{QF}}$ with the instantiation $\bodyform[j/i]$
\emph{if} 
that offset appears in the concrete offsets of a violated quantifier-free clause (or a violated instantiation).
Rather than computing from scratch a new model for the strengthened query,
we fix the values of all the array cells (and variables) according to their interpretation in $m_d$
except for the arrays that are accessed with $\boundvar$,
those for which a new interpretation is sought.
If the resulting query has a model, we apply assignment duplication on it.
This time, to avoid overwriting,
the duplication is not applied to the offsets involved in violations.

\begin{example}\label{ex:repair}
Continuing \Cref{ex:duplicate},
the violated clause in the model $m_d$ is $s[k - 1] = 8$, and its concrete access is $s[6]$.
The concrete access in the instantiation $(s[i - 1] \neq 0)[7/i]$ that was omitted in $\varphi_{\mathit{QF}}$ is also $s[6]$,
so we add it to $\varphi_{\mathit{QF}}$.
In addition, we concretize the values of $n$ and $k$ according to $m_d$.
The resulting strengthened query and its possible model are:
\[
\begin{array}{c}
\varphi_{\mathit{QF}} \land (s[6] \neq 0) \land (n = 7) \land (k = 7) \\
\{ n \mapsto 7, k \mapsto 7, s \mapsto [1,0,0,0,0,0,8,0] \}
\end{array}
\]

Then, we duplicate again,
but this time while skipping over the cell $s[6]$.
Similarly to the first duplication, $v$ is set to 1,
but the value of $s[j]$ is updated only for $1 \leq j \leq 5$,
thus avoiding the original violation.
The resulting model indeed  satisfies $\varphi$:
\[ m_r \triangleq \{ n \mapsto 7, k \mapsto 7, s \mapsto [1,1,1,1,1,1,8,0] \} \]
\end{example}
%

\textbf{\textit{(4) Fallback.}}
If no model $m_r$ is found, or if it does not satisfy $\varphi$,
we ask the SMT solver to find a model for $\varphi$.

\pbreak

\section{Implementation}\label{sec:implementation}


We implemented our state merging approach on top of the KLEE~\cite{klee} symbolic execution engine,
configured with LLVM 7.0.0~\cite{llvm}.
Our approach generates quantified queries over arrays and bit vectors,
so we use Z3~\cite{z3} (version 4.8.17) as the underlying SMT solver. 
We extended \klee's expression language to support quantified formulas,
and modified some parts of the solver chain accordingly.
We implemented our solving procedure (\Cref{sec:solving})
as an additional component in the solver chain.
To implement the \textit{hash} function used by the pattern-based state merging approach (\Cref{sec:pattern-based}),
we relied on the expression hashing utility of \klee
and modified it by assigning a pre-defined hash value to all constants.
To extract the regular patterns from the execution trees,
we used a basic regular expression matching algorithm.
If our hash function is not valid for a given generated execution tree (\Cref{def:hash}),
or the number of extracted regular patterns in that tree exceeds a user-specified threshold,
then we fallback to standard state merging (\Cref{def:state-merging}).
Our implementation is available at~\cite{smq-implementation}.

\pbreak

\section{Evaluation} \label{sec:evaluation}

\revision{Evaluating a state-merging approach requires determining the desired merging points,
\ie the code segments where state merging should be applied.
In our case,
this translates to identifying code segments that produce merging operations that involve many symbolic states.
To do so,
we evaluate our approach in the context of the \textit{symbolic-size} memory model~\cite{symsize-model}.
This model supports bounded symbolic-size objects, \ie objects whose size can have a range of values,
limited by a user-specified \emph{capacity} bound.\footnote{This is in contrast to the standard \textit{concrete-size} model where every object has a concrete size.}
It was observed in~\cite{symsize-model} that loops operating on symbolic-size objects typically produce many symbolic states,
and state-merging was suggested to combat the ensued state explosion problem.
Thus, this memory model provides a suitable basis for evaluating our state-merging approach.
Furthermore, the automatic detection of merging points in~\cite{symsize-model} avoids the need for manual annotations.
We emphasize, however, that our technique is independent of the symbolic-size memory model itself  (see~\Cref{sec:threats}).
That said,
the symbolic-size memory model does have the potential to produce more challenging merging operations than the concrete-size model
as it considers a larger state space.}

The following modes are the main subjects of comparison:
The \patternabv mode is the pattern-based state merging approach described in \Cref{sec:pattern-based}
which partitions the symbolic states into merging groups based on regular patterns in the execution tree,
and uses quantifiers to encode the merged path constraints.
In the \patternabv mode,
the incremental state merging approach (\Cref{sec:incremental}) and the solving procedure (\Cref{sec:solving}) are enabled.
The \cfg mode is the state merging approach discussed above (\smopt mode from~\cite{symsize-model}),
which partitions the symbolic states into merging groups according to their exit point from the loop in the CFG,
and uses the standard \qfabv encoding~\cite{stp} (disjunctions and \ite expressions).
The \base mode is the forking approach used in vanilla \klee~\cite{klee}.

The following research questions guide our evaluation:
\begin{enumerate}
\item[(RQ1)] Does \patternabv improve standard state merging (\cfg)?
\item[(RQ2)] Does \patternabv improve standard symbolic execution (\base)?
\item[(RQ3)] What is the significance of each component of \patternabv?
\end{enumerate}

\subsection{Benchmarks}
The benchmarks used in our evaluation are listed in~\Cref{table:benchmarks}.
These benchmarks were chosen as they are challenging for symbolic execution and provide numerous opportunities for applying state merging.
\revision{In each benchmark, we analyzed a set of subjects (APIs and whole programs)
whose inputs (parameters, command-line arguments, \etc) can be modeled using symbolic-size objects, \ie arrays and strings.}
In \libosip~\cite{libosip}, \libtasn~\cite{libtasn1}, and \libpng~\cite{libpng},
the test drivers for the APIs were taken from~\cite{symsize-model}.\footnote{
\revision{We noticed that some of the APIs from \libosip that were used in~\cite{symsize-model} are similar,
\ie different APIs with the same internal functionality.
The analysis of such APIs leads to the same results, therefore, we excluded them from the evaluation to avoid redundancy.}}
In \wget~\cite{wget}, a library for retrieving files using widely used internet protocols (HTTP, \etc),
we reused the test drivers from the existing fuzzing test suite whenever possible,
and for other APIs, we constructed the test drivers manually.
\revision{In \apr~\cite{apr} (Apache Portable Runtime), a library that provides a platform-independent abstraction of operating system functionalities,
we constructed test drivers for APIs from several modules (\textit{strings}, \textit{file\_io} and \textit{tables})
which manipulate strings, file-system paths, and data structures.
In \jsonc~\cite{jsonc}, a library for decoding and encoding JSON objects,
we constructed test drivers for APIs that manipulate string objects.
In \busybox~\cite{busybox}, a software suite that provides a collection of Unix utilities,
we focused on utilities whose input comes from command-line arguments and files,
which can be symbolically modeled using \klee's \textit{posix} runtime.
We did not analyze utilities whose behavior depends on the state of system resources
(process information, permissions, file-system directories, etc.),
since \klee has no symbolic modeling for those.
To prevent the symbolic executor from getting stuck in \texttt{getopt()},
the routine used in \busybox to parse command line arguments,
we added the restriction that symbolic command line arguments do not begin with a `\texttt{-}' character.}


\subsection{Setup}
We run every mode under the symbolic-size memory model~\cite{symsize-model} with the following configuration:
a DFS search heuristic, a one-hour time limit, and a 4GB memory limit.
The capacity settings in each of the benchmarks are shown in~\Cref{table:benchmarks}.\footnote{
In \libosip, \libtasn, and \libpng,
the capacity settings were set similarly to the experiments from~\cite{symsize-model}.
}

In every experiment, we use the following metrics to compare between the modes:
analysis time and line coverage computed with GCov~\cite{gnugcov}.
When the compared modes have the same exploration order, 
we additionally use the path coverage metric, \ie the number of explored paths.

Each benchmark consists of multiple subjects,
so when comparing the two modes,
we measure the relative speedup and the relative increase in coverage for each subject.
Note that when we measure the average (and median) speedup, for example,
the speedup in the subjects where both modes timed out is always 1\x.
Similarly, when we measure coverage,
the coverage in the subjects where both modes terminated, \ie completed the analysis before hitting the timeout, is always identical.
To separate the subjects where the results are trivially identical,
we report the average (and median) over a \emph{subset} of the subjects depending on the evaluated metric:
When measuring analysis time,
we consider the subset of the subjects where at least one of the modes terminated.
When measuring coverage,
we consider the subset of the subjects where at least one of the modes timed out.
In~\citeapp[C.1.1],
we additionally report the average (and median) when computed over all the subjects.

We ran our experiments on several machines (Intel i7-6700 @ 3.40GHz with 32GB RAM) with Ubuntu 20.04.

\begin{table}[t]
\caption{Benchmarks.}
\label{table:benchmarks}
\centering
{\footnotesize
\begin{tabular}{|l|l|r|l|l|}
\cline{1-5}
\multicolumn{1}{|c|}{} &
\multicolumn{1}{c|}{\text{Version}} &
\multicolumn{1}{c|}{\text{SLOC}} &
\multicolumn{1}{c|}{\text{\#Subjects}} &
\multicolumn{1}{c|}{\text{Capacity}} \\
\hline
\libosip & 5.2.1 & 18,783 & 35 & 10 \\ \hline
\wget & 1.21.2 & 100,785 & 31 & 200 \\ \hline
\libtasn & 4.16.0 & 15,291 & 13 & 100 \\ \hline
\libpng & 1.6.37 & 56,936 & 12 & 200 \\ \hline
\revisionrow \apr & 1.6.3 & 60,034 & 20 & 50 \\ \hline
\revisionrow \jsonc & 0.15 & 8,167 & 5 & 100 \\ \hline
\revisionrow \busybox & 1.36.0 & 198,500 & 30 & 100 \\ \hline
\end{tabular}
}
\end{table}

\subsection{Results: \patternabv vs. \cfg}
\label{eval:rq1}

\begin{figure*}
    \centering
    \begin{subfigure}[b]{0.30\textwidth}
        \centering
        \includegraphics[width=\textwidth]{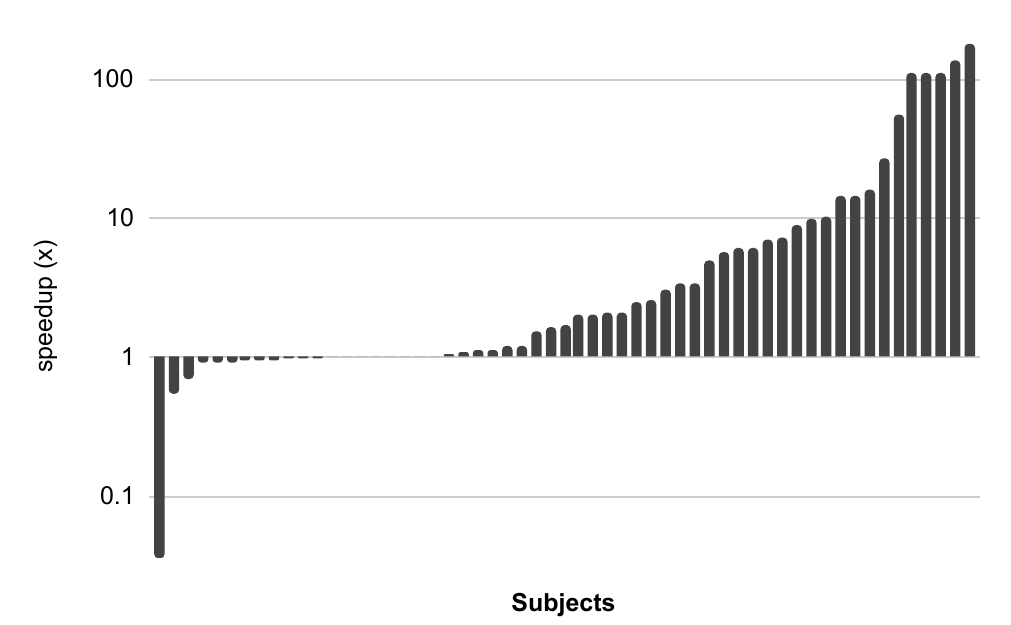}
        \caption{Speedup in analysis time ($\times$) in the subjects where at least one of the modes terminated (in log-scale).}
        \label{fig:cfg-speedup}
    \end{subfigure}
    \hfill
    \begin{subfigure}[b]{0.30\textwidth}
        \centering
        \includegraphics[width=\textwidth]{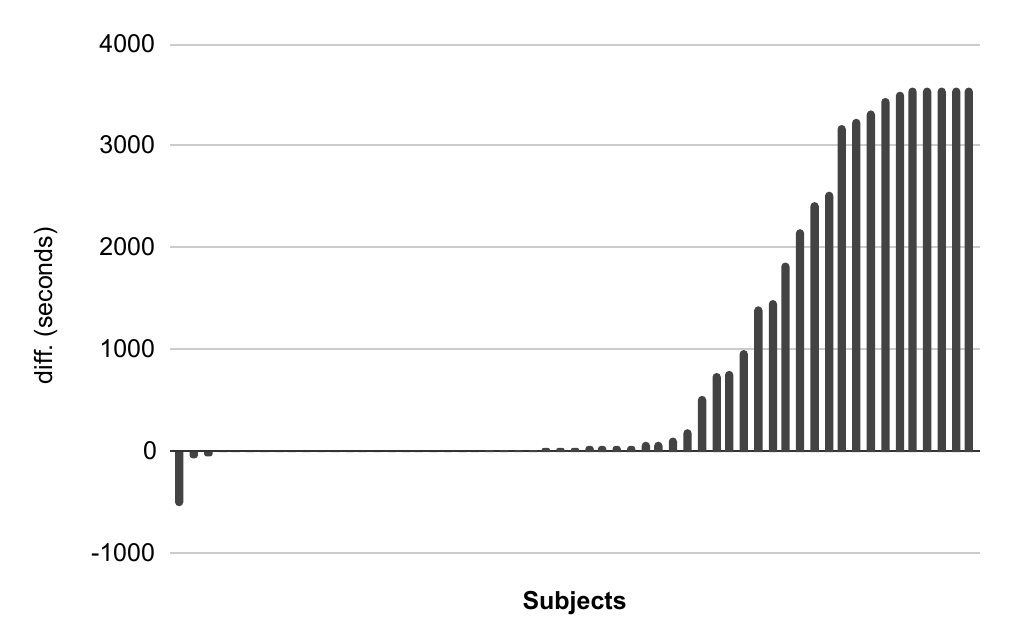}
        \caption{Difference in analysis time (\emph{seconds}) in the subjects where at least one of the modes terminated.}
        \label{fig:cfg-diff}
    \end{subfigure}
    \hfill
    \begin{subfigure}[b]{0.30\textwidth}
        \centering
        \includegraphics[width=\textwidth]{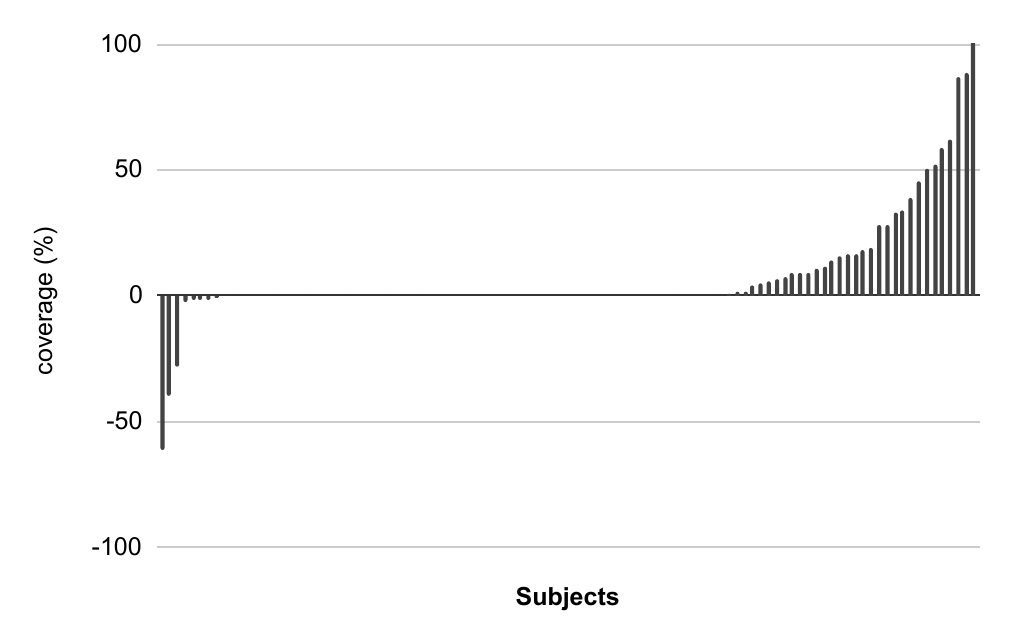}
        \caption{Relative increase in coverage (\%) in the subjects where at least one of the modes timed out.}
        \label{fig:cfg-coverage}
    \end{subfigure}
    \caption{Breakdown of the improvement of \patternabv over \cfg per subject.}
    \label{fig:cfg-breakdown}
\end{figure*}

In this experiment,
we compare between the performance of the state merging modes: \patternabv and \cfg.
The results are shown in \Cref{table:vs-cfg} and \Cref{fig:cfg-breakdown}.

\paragraph{Analysis Time}
Column \textit{Speedup} in \Cref{table:vs-cfg} shows the (average, median, minimum, and maximum) speedup of \patternabv compared to \cfg
in the subjects where at least one of the modes terminated.
\revision{Column \textit{\#} shows the number of considered subjects out of the total number of subjects.}
\revision{In \libosip, \wget, \apr, \jsonc, and \busybox,
\patternabv was significantly faster in many subjects,
and in \libtasn and \libpng,
the analysis times were roughly identical.}
\Cref{fig:cfg-speedup} breaks down the speedup of \patternabv compared to \cfg per subject.
\revision{Overall, there were 12 subjects where \patternabv was slower than \cfg.}
In \libosip, \patternabv was slower only in one API.
In this case,
the slowdown of 0.03\x (from 20 to 554 seconds) was caused by a small number of queries (9) that our solving procedure (\Cref{sec:solving}) failed to solve,
and whose solving using the SMT solver required most of the analysis time.
In \wget,
\patternabv was slower in two APIs.
In one case, the slowdown was caused by the computational overhead of the incremental state merging approach.
In the other case, the slowdown was caused by a relatively high number of queries that our solving procedure failed to solve.
In \libtasn, \patternabv was slower in seven APIs,
but the time difference in these cases was rather minor (roughly 10 seconds).
In \libpng, \patternabv was slightly slower in one API due to the computational overhead of extracting regular patterns.
\revision{In \busybox, \patternabv was slower in one utility with a minor time difference of two seconds.}
Column \textit{Diff.} in \Cref{table:vs-cfg} shows the difference between \patternabv and \cfg in terms of the total time required to analyze all the subjects.
Note that the time difference is interpreted as zero in subjects where both modes are timed out. 
\revision{In \libosip, \wget, \apr, and \busybox,
\patternabv achieved a considerable reduction of roughly 8, 4, 1, and 3 hours, respectively.
In \jsonc, \patternabv achieved a reduction of roughly 20 minutes,
and in \libtasn and \libpng, the time difference was minor.}
\Cref{fig:cfg-diff} breaks down the time difference between \patternabv and \cfg per subject.

\paragraph{Coverage}
Column \textit{Coverage} in \Cref{table:vs-cfg} shows the (average, median, minimum, and maximum) relative increase in line coverage of \patternabv over \cfg
in the subjects where at least one of the modes timed out.
\revision{Again, column \textit{\#} shows the number of considered subjects.}
In \libosip and \wget,
\patternabv achieved higher coverage in many cases.
In \libtasn,
\patternabv resorted to standard state merging in most cases,
as it did not find regular (and formula) patterns.
Therefore, the results were similar to those of \cfg, and coverage was not improved.
In \libpng, the coverage was roughly identical in all the APIs except for two APIs where \patternabv achieved an improvement of 8.69\% and 18.33\%.
\revision{In \apr, the coverage was identical in all the APIs except for two cases
where \patternabv had an increase of 16.62\% and a decrease of 2.12\%.
In \jsonc,
there was only one API where one of the modes timed out,
and in this case, \cfg achieved higher coverage.
In \busybox,
there were 23 cases where at least one of the modes timed out.
In four cases, \patternabv achieved an improvement of 3.98\%-15.45\%,
and in two cases, \cfg achieved an improvement of 1.15\% and 61.78\%.
In the remaining 17 cases, the coverage was identical.
(In most of these cases, \patternabv did not find formula patterns, resulting in identical explorations.)}
Column \textit{Diff.} in \Cref{table:vs-cfg} shows the difference between \patternabv and \cfg in terms of the total number of covered lines across all the subjects.
Again, note that there is no difference in coverage in subjects where both modes terminated.
\revision{It is possible to have an improvement in average coverage but not in total line difference (\apr) and vice versa (\busybox).
This happens due to shared code that is covered by only one mode in one subject
but covered by the other mode in other subjects.}
\Cref{fig:cfg-coverage} breaks down the coverage improvement of \patternabv over \cfg per subject.

\begin{table*}[t]
\caption{Comparison of \patternabv vs. \cfg.}
\label{table:vs-cfg}
\centering
{\footnotesize
\begin{tabular}{|l|r|r|r|r|r|r|r|r|r|r|r|r|}
\cline{1-13}
\multicolumn{1}{|c|}{} &
\multicolumn{6}{c|}{\text{Time}} &
\multicolumn{6}{c|}{\text{Coverage (\%)}} \\

\multicolumn{1}{|c|}{} &
\multicolumn{5}{c|}{\text{Speedup ($\times$)}} &
\multicolumn{1}{c|}{\text{Diff. (seconds)}} &
\multicolumn{5}{c|}{\text{}} &
\multicolumn{1}{c|}{\text{Diff. (lines)}} \\

\multicolumn{1}{|c|}{} &
\multicolumn{1}{c|}{\#} & \multicolumn{1}{c|}{Avg.} & \multicolumn{1}{c|}{Med.} & \multicolumn{1}{c|}{Min.} & \multicolumn{1}{c|}{Max.} &
\multicolumn{1}{c|}{} &
\multicolumn{1}{c|}{\#} & \multicolumn{1}{c|}{Avg.} & \multicolumn{1}{c|}{Med.} & \multicolumn{1}{c|}{Min.} & \multicolumn{1}{c|}{Max.} &
\multicolumn{1}{c|}{} \\

\hline
\libosip & 16/35 & 7.18 & 5.50 & 0.03 & 180.00 & 27668 & 28/35 & 20.45 & 9.00 & 0.00 & 88.63 & 291 \\
\hline
\wget & 11/31 & 2.69 & 1.67 & 0.54 & 14.69 & 12942 & 24/31 & 15.02 & 0.00 & -40.00 & 300.00 & 89 \\
\hline
\libtasn & 7/13 & 0.94 & 0.95 & 0.90 & 0.96 & -41 & 6/13 & 0.00 & 0.00 & 0.00 & 0.00 & 0 \\
\hline
\libpng & 1/12 & 0.70 & 0.70 & 0.70 & 0.70 & -9 & 11/12 & 2.03 & 0.00 & -2.88 & 18.33 & 104 \\
\hline
\revisionrow \apr & 10/20 & 3.50 & 1.63 & 1.00 & 138.46 & 4375 & 11/20 & 1.31 & 0.00 & -2.12 & 16.62 & 0 \\
\hline
\revisionrow \jsonc & 4/5 & 3.16 & 2.97 & 2.00 & 5.76 & 1149 & 1/5 & 0.81 & 0.81 & 0.81 & 0.81 & 1 \\
\hline
\revisionrow \busybox & 8/30 & 1.68 & 1.07 & 0.92 & 16.20 & 10100 & 23/30 & -1.08 & 0.00 & -61.78 & 15.45 & 74 \\
\hline
\end{tabular}
}
\end{table*}

\paragraph{Scaling}
The main obstacle in applying state merging originates from the introduction of disjunctive constraints and \ite expressions,
especially when the number of states to be merged is high.
\revision{We evaluate the ability of our approach to cope with a particular aspect of this challenge where the states are generated by loops iterating over large data objects, a frequent situation in our experience.}
Technically, we conducted a case study on \libosip, one of our benchmarks,
where we gradually increase the \textit{capacity} of symbolic-size objects.
When the capacity is increased,
the size of the symbolic-size objects is potentially increased as well.
This typically leads to additional forks, for example, in loops that operate on symbolic-size objects. 
As we apply state merging in such loops,
this eventually results in more complex merging operations.
Thus, increasing the capacity allows us to measure how each mode scales w.r.t. the number of merged states.
In this experiment,
we run each API in each of the state merging modes (\patternabv and \cfg) under several different capacity settings.
The results are shown in \Cref{table:scale}.

As can be seen,
\patternabv achieved better results than \cfg in all the capacity settings.
In general, when the capacity is increased, there are typically more forks and queries,
which makes the analysis of size-dependent loops harder for both modes.
Therefore, the coverage improvement was less significant under the highest capacity settings (100 and 200) compared to the lower capacity settings.
Note also that under those capacity settings, there were only five APIs in which at least one of the modes terminated.
We observed that in these APIs, the analysis time increased in both modes when the capacity was increased.
However, with \cfg, the analysis time increased more significantly,
so the speedup under the highest capacity setting (200) was greater.
\revision{This indicates that our approach is less sensitive to the input capacity
and hence to the resulting number of merged states.}

\rqanswer{1}{
\patternabv outperforms \cfg in many cases
and scales better in executing complex state merging operations.
}

\begin{table}[t]
\caption{Comparison of \patternabv vs. \cfg under different capacity settings (column \textit{Capacity}) in \libosip.}
\label{table:scale}
\centering
{\footnotesize
\begin{tabular}{|l|r|r|r|r|r|r|}
\cline{1-7}
\multicolumn{1}{|c|}{\text{Capacity}} &
\multicolumn{3}{c|}{\text{Speedup ($\times$)}} &
\multicolumn{3}{c|}{\text{Coverage (\%)}} \\
\multicolumn{1}{|c|}{} &
\multicolumn{1}{c|}{\#} & \multicolumn{1}{c|}{Avg.} & \multicolumn{1}{c|}{Med.} &
\multicolumn{1}{c|}{\#} & \multicolumn{1}{c|}{Avg.} & \multicolumn{1}{c|}{Med.} \\
\hline
10 & 16/35 & 7.18 & 5.50 & 28/35 & 20.45 & 9.00 \\
\hline
20 & 13/35 & 4.58 & 5.53 & 29/35 & 23.41 & 19.29 \\
\hline
50 & 12/35 & 1.99 & 2.43 & 30/35 & 15.19 & 10.63 \\
\hline
100 & 5/35 & 2.99 & 2.75 & 30/35 & 10.23 & 2.32 \\
\hline
200 & 5/35 & 4.81 & 6.11 & 30/35 & 4.22 & 0.00 \\
\hline
\end{tabular}
}
\end{table}

\subsection{Results: \patternabv vs. \base}
\label{eval:rq2}

In this experiment,
we compare the performance of \patternabv and \base,
\ie standard symbolic execution that uses the forking approach.
The results are shown in \Cref{table:vs-base}.

Column \textit{Speedup} shows the (average and median) speedup of \patternabv compared to \base
in the subjects where at least one of the modes terminated.
\revision{
As can be seen,
\patternabv achieved a considerable speedup in the majority of the benchmarks.
Overall, there were nine subjects in which \patternabv was slower than \base.
In three of these cases,
the time difference was minor (roughly 5 seconds).
In the other cases,
the slowdown was caused by the computational overhead of the incremental state merging approach
and the complex constraints that were introduced during the state merging.
Regarding timeouts,
there were 20 subjects in which \base timed out and \patternabv terminated,
and only one subject in which \patternabv timed out and \base terminated.}

Column \textit{Coverage} shows the (average and median) relative increase in line coverage of \patternabv over \base
in the subjects where at least one of the modes timed out.
\patternabv achieved higher coverage in many subjects,
especially in \libosip and \libpng.
\revision{In most of the cases in \libtasn, \apr, and \jsonc,
both modes covered most of the reachable lines in a relatively early stage,
so the coverage was similar.
In \wget and \busybox,
\patternabv achieved higher coverage in some of the cases,
but there were also cases in which \base achieved higher coverage.}
In general,
this is a consequence of the known tradeoff between forking and state merging:
The forking approach explores more paths but generates less complex constraints.

In addition, we observed that there were four subjects in which \base ran out of memory.
In two of these cases, \base finished the analysis before \patternabv,
but its analysis was incomplete since \klee prunes the search space once the memory limit is reached.

For space reasons,
the breakdown of the improvement of \patternabv over \base per subject is shown in~\citeapp[C.1.2].

\rqanswer{2}{
\patternabv outperforms \base in many cases,
however, the known tradeoff between state merging and forking remains.
}

\begin{table}[t]
\caption{Comparison of \patternabv vs. \base.}
\label{table:vs-base}
\centering
{\footnotesize
\begin{tabular}{|l|r|r|r|r|r|r|}
\cline{1-7}
\multicolumn{1}{|c|}{} &
\multicolumn{3}{c|}{\text{Speedup ($\times$)}} &
\multicolumn{3}{c|}{\text{Coverage (\%)}} \\
\multicolumn{1}{|c|}{} &
\multicolumn{1}{c|}{\#} & \multicolumn{1}{c|}{Avg.} & \multicolumn{1}{c|}{Med.} &
\multicolumn{1}{c|}{\#} & \multicolumn{1}{c|}{Avg.} & \multicolumn{1}{c|}{Med.} \\

\hline
\libosip & 17/35 & 11.21 & 3.10 & 28/35 & 11.43 & 1.88 \\
\hline
\wget & 12/31 & 2.75 & 3.72 & 24/31 & -2.32 & 0.00 \\
\hline
\libtasn & 7/13 & 4.94 & 9.30 & 7/13 & 1.49 & 0.00 \\
\hline
\libpng & 1/12 & 2.46 & 2.46 & 11/12 & 23.59 & 7.14 \\
\hline
\revisionrow \apr & 10/20 & 8.40 & 3.91 & 14/20 & -0.15 & 0.00 \\
\hline
\revisionrow \jsonc & 4/5 & 1.36 & 3.09 & 2/5 & 0.82 & 0.82 \\
\hline
\revisionrow \busybox & 9/30 & 2.43 & 2.51 & 22/30 & -2.76 & 0.00 \\
\hline
\end{tabular}
}
\end{table}

\subsection{Results: Component Breakdown}
\label{eval:rq3}

Now, we evaluate the significance of the components used in our pattern-based state merging approach (\ie \patternabv).

\subsubsection{\textbf{Solving Procedure}}
\label{eval:solving-procedure}

To evaluate our solving procedure (\Cref{sec:solving}),
we ran each subject in two versions of \patternabv:
one that relies only on the SMT solver (vanilla Z3) and another one that uses our solving procedure.
Both modes are run with the incremental state merging approach enabled.

\revision{To evaluate the impact of the solving procedure,
we show in \Cref{table:solving-procedure} its effect on analysis time and coverage in the relevant subsets.
Here, the two modes have the same exploration order,  
so we use the path coverage metric as well.
In \libosip, \wget, \apr, \jsonc, and \busybox,
our solving procedure generally leads to lower analysis times
and higher (line or path) coverage.
The results were mostly similar in \libtasn and \libpng since the number of quantified queries was relatively low.
The only exception was one of the APIs in \libpng, where the path coverage was increased by 39.51\%.}

\begin{table}[t]
\caption{Impact of solving procedure.}
\label{table:solving-procedure}
\centering
{\footnotesize
\begin{tabular}{|l|r|r|r|r|r|r|r|r|}
\cline{1-9}
\multicolumn{1}{|c|}{} &
\multicolumn{3}{c|}{\text{Speedup ($\times$)}} &
\multicolumn{5}{c|}{\text{Coverage (\%)}} \\
\multicolumn{1}{|c|}{\text{}} &
\multicolumn{3}{c|}{} &
\multicolumn{1}{c|}{} &
\multicolumn{2}{c|}{\text{Line}} &
\multicolumn{2}{c|}{\text{Path}} \\

\multicolumn{1}{|c|}{} &
\multicolumn{1}{c|}{\#} & \multicolumn{1}{c|}{Avg.} & \multicolumn{1}{c|}{Med.} &
\multicolumn{1}{c|}{\#} & \multicolumn{1}{c|}{Avg.} & \multicolumn{1}{c|}{Med.} & \multicolumn{1}{c|}{Avg.} & \multicolumn{1}{c|}{Med.} \\

\hline
\libosip & 16/35 & 1.55 & 1.57 & 19/35 & 0.26 & 0.00 & 89.31 & 72.82 \\
\hline
\wget & 11/31 & 4.28 & 3.62 & 27/31 & 14.81 & 0.00 & 110.17 & 30.94 \\
\hline
\libtasn & 7/13 & 0.99 & 0.99 & 6/13 & 0.00 & 0.00 & -0.74 & -0.24 \\
\hline
\libpng & 1/12 & 1.03 & 1.03 & 11/12 & -0.23 & 0.00 & 2.62 & 0.00 \\
\hline
\revisionrow \apr & 10/20 & 2.86 & 3.49 & 10/20 & 0.00 & 0.00 & 38.31 & 5.57 \\
\hline
\revisionrow \jsonc & 4/5 & 2.89 & 2.33 & 1/5 & 0.00 & 0.00 & 79.49 & 79.49 \\
\hline
\revisionrow \busybox & 8/30 & 1.29 & 1.09 & 23/30 & 0.52 & 0.00 & 9.53 & 1.65 \\
\hline
\end{tabular}
}
\end{table}

\subsubsection{\textbf{Incremental State Merging}}
\label{eval:incremental}

To evaluate the incremental state merging approach (Section~\ref{sec:incremental}),
we run each subject in two versions of \patternabv:
one that disables incremental state merging and another one that enables it.
The results are shown in \Cref{table:incremental}.

In \libosip,
there were relatively many loops where incremental state merging was successfully applied, \ie reduced the number of explored paths.
This resulted in a significant speedup and higher line coverage.
In \wget,
there were four APIs where incremental state merging could be applied,
and in two of these cases, the coverage was improved by 33.33\% and 300.00\%.
\revision{In \apr,
there were four APIs where incremental state merging could be applied,
and in one of these cases, the analysis time was reduced by 138.46\x and the coverage was improved by 16.62\%.
In \busybox,
there were two utilities where incremental state merging could be applied,
and in these cases, the coverage was improved by 11.33\% and 15.45\%.
In \libtasn, \libpng, and \jsonc,
there were no loops where incremental state merging could be applied.
In some cases, this resulted in a minor performance penalty due to the computational overhead of the approach,
which mainly comes from the need to maintain the snapshots of the non-active symbolic states in the execution tree.}


\rqanswer{3}{
All the components contribute to the performance of \patternabv.
}

\begin{table}[t]
\caption{Impact of incremental state merging.}
\label{table:incremental}
\centering
{\footnotesize
\begin{tabular}{|l|r|r|r|r|r|r|}
\cline{1-7}
\multicolumn{1}{|c|}{} &
\multicolumn{3}{c|}{\text{Speedup ($\times$)}} &
\multicolumn{3}{c|}{\text{Coverage (\%)}} \\
\multicolumn{1}{|c|}{} &
\multicolumn{1}{c|}{\#} & \multicolumn{1}{c|}{Avg.} & \multicolumn{1}{c|}{Med.} &
\multicolumn{1}{c|}{\#} & \multicolumn{1}{c|}{Avg.} & \multicolumn{1}{c|}{Med.} \\

\hline
\libosip & 16/35 & 6.78 & 2.80 & 28/35 & 18.98 & 5.83 \\
\hline
\wget & 11/31 & 0.97 & 0.97 & 20/31 & 16.66 & 0.00 \\
\hline
\libtasn & 7/13 & 0.96 & 0.98 & 6/13 & 0.00 & 0.00 \\
\hline
\libpng & 1/12 & 0.96 & 0.96 & 11/12 & 2.35 & 0.00 \\
\hline
\revisionrow \apr & 11/20 & 1.60 & 1.00 & 11/20 & 1.71 & 0.00 \\
\hline
\revisionrow \jsonc & 4/5 & 1.01 & 1.01 & 1/5 & 0.00 & 0.00 \\
\hline
\revisionrow \busybox & 8/30 & 0.98 & 1.00 & 20/30 & 0.76 & 0.00 \\
\hline
\end{tabular}
}
\end{table}

\revision{%
\subsection{Found Bugs}
We found two bugs during our experiments with \busybox.
In both cases,
a null-pointer dereference occurred in the implementation of \code{realpath} in \kleeuclibc,
\klee's modified version of \uclibc~\cite{uclibc}.
We reported the bugs, which were confirmed and fixed by the official maintainers.\footnote{
\url{https://github.com/klee/klee-uclibc/pull/47}
}
We note that these bugs were detected by \patternabv and \base,
but were not found by \cfg due to a timeout.}

\subsection{Threats to Validity}\label{sec:threats}
First, our implementation may have bugs.
To validate its correctness,
we performed a separate experiment where each subject was run in the \patternabv mode with a timeout of one hour.
During these runs,
we validated that every executed state merging operation is correct w.r.t. \Cref{theorem:state-merging}.
In addition, for every query that our solving procedure was able to solve,
we validated the consistency of the reported result w.r.t. the underlying SMT solver.

%
Second, our choice of benchmarks might not be representative enough.
That said,
we chose a diverse set of real-world benchmarks used in prior work~\cite{loops:pldi19,symsize-model,kdalloc:ecoop22}.
In addition,
we used benchmarks that process inputs of both binary and textual formats.

\revision{Third, we evaluated our approach in the context of the symbolic-size model~\cite{symsize-model}.
To address the threat that our approach may be beneficial only in the context of that particular memory model,
we performed an additional experiment using the standard concrete-size memory model.
In this experiment,
we set the concrete sizes of the input objects according to the capacity configuration in \Cref{table:benchmarks},
and apply state merging in loops whose conditions depend on these sizes, as we do in our original experiments.
The results, shown in~\citeapp[C.3], lead to conclusions similar to the ones drawn from the original experiments.}

\revision{Fourth, the search heuristic might affect the coverage when the exploration does not terminate.
To address the threat that our results may be valid only for the DFS search heuristic,
we performed an additional experiment using the default search heuristic in KLEE.
The results, shown in~\citeapp[C.4], are comparable.
}

\subsection{Discussion}

Taking a high-level view of the experiments,
we observe that our approach brings significant gains w.r.t. both baselines in most of the benchmarks
(\libosip, \wget, \revision{\apr, \jsonc, and \busybox}).
This is because these benchmarks contain an abundant number of size-dependent loops
that generate expressions that are linearly dependent on the number of repetitive parts in the path constraints,
which leads to the detection of many regular (and formula) patterns.
In \libtasn and \libpng, however, 
most of the size-dependent loops generate expressions that cannot be synthesized with our approach,
for example, aggregate values such as the sum of array contents.
As a result, relatively few formula patterns are detected.
Nevertheless, in these cases,
our approach still preserves the benefits of standard state merging w.r.t. standard symbolic execution.


\pbreak

\section{Related Work}\label{sec:related}

Compact symbolic execution~\cite{compact-s-e} uses quantifiers to encode the path conditions of cyclic paths
that follow the \emph{same} control flow path in each iteration and
update all the variables in a regular manner.
This allows them to encode the effect of unbounded repetitions of \emph{some} of the cyclic paths in the program.
In contrast, we seek regularity at the level of the constraints and, therefore, do not rely on uniformity in the control flow graph.
In \introfunc (\Cref{fig:example}), for example, they can only summarize the paths in which
either all the characters of \code{s} are matched with the first character of \code{chars} (the then branch) or the first character of \code{s} is unmatched (the else branch).
In contrast, our approach can summarize \emph{all} paths up to a given bound using two merged states.
Furthermore, \cite{compact-s-e} solves quantified queries using a standard solver as opposed to our specialized solving procedure.

Loop-extended symbolic execution~\cite{loop-symbex}  summarizes input-dependent loops.
It uses static analysis to infer linear relations between variables and trip count variables tracking the number of iterations in the loop.
Godefroid et al.~\cite{sageLoopsummary} propose a dynamic approach 
for inferring invariants in input-dependent loops, which allows them to partially summarize the loop's effect on induction variables.
In contrast, our approach does not rely on induction variables or the number of loop iterations. 
Kapus et al.~\cite{loops:pldi19} summarize string loops by synthesizing calls to standard string functions. 
S-Looper~\cite{slooper} introduces string constraints that can be solved
by solvers that support the string theory. 
Our approach is not restricted to string loops and does not require a solver supporting string theory.

%
%
%

%
%
%
%
%
%

Veritesting~\cite{veritesting:icse14} improves the performance of symbolic execution by merging similar execution paths.
Given a symbolic branch,
veritesting summarizes side effects from both branch sides to avoid path explosion.
%
Java Ranger~\cite{java-ranger} extends veritesting of Java programs to support dynamically dispatched methods,
by using the runtime information available during the analysis.
MultiSE~\cite{multise} summarizes updates to values by efficiently guarding each value with a path predicate. 
Kuznetsov et al.~\cite{merging:pldi12} merge symbolic states based on a
query count heuristic that estimates if the merging would reduce the solving time in the future.
%
%
Trabish et al.~\cite{symsize-model} perform state merging in loops that depend on objects whose size is symbolic.
They reduce the size of the encoding in the resulting merged states using the execution tree,
but still rely on disjunctions and \ite expressions, therefore unable to achieve the reduction obtained with our approach.
We explicitly compared our technique with theirs (referred to as \cfg in \Cref{sec:evaluation})
and show that our approach performs better in many cases.
%
The works mentioned above do not address the encoding explosion problem caused by using disjunctions and \ite expressions.

\revision{There are many works on handling quantified formulas~\cite{reynolds2013quantifier,detlefs2005simplify,reynolds2014finding,CAV09mbqi,reynolds2018revisiting,ematching,bansal2015deciding,bradley2006s}.
Our solving procedure (\Cref{sec:solving}) is mainly designed to solve satisfiable queries.
It adapts ideas from E-matching~\cite{ematching} and model-based quantifier instantiation~\cite{CAV09mbqi} to our specific needs.}

\pbreak

\section{Conclusions and Future Work} \label{sec:conclusion}

\nrx{Merge two paragraphs of conclusions.}

We propose a state merging approach that significantly reduces the encoding complexity of merged symbolic states and
show through our evaluation that this is a promising direction toward scaling state merging in symbolic execution.

Our approach automatically detects regular patterns to partition similar symbolic states into merging groups.
For each group, we synthesize a formula pattern that enables an efficient encoding of the merged symbolic state using quantifiers.
Extracting more complex patterns, \eg beyond linear formulas, can further improve the applicability of our approach.




\paragraph{Acknowledgements} 
This research was partially funded by the Israel Science Foundation (ISF)
grants No. 1996/18 and No. 1810/18 and by Len Blavatnik and the Blavatnik Family foundation.



\newpage

\bibliography{bib_files/cadar-macros,bib_files/cadar,bib_files/cadar-crossrefs,string-long,biblio,klee,combined}

\appendix 


\section{Proofs}

\subsection{Proof of Theorem 3.7}
Suppose that the execution tree of the merged code fragment is $t$ (with root $r$).
According to the validity of $t$ (Section 2), the following holds for every $j = 1, ..., n$:
\[ n_j.s.pc \equiv r.s.pc \land \treepc(n_j) \]
Without loss of generality, we assume that $r.s.pc \equiv true$.
According to Definition 2.1:
\[ s'.pc \triangleq \bigvee_{j = 1}^{n}\treepc(n_j) \]
We assumed that $\{(n_j, k_j)\}_{j = 1}^{n}$ match the formula pattern $(\varphi_1, \varphi_2(x), \varphi_3(x))$, so:
\[ \treepc(n_j) \eqsyn \varphi_1 \land \big( \bigwedge_{i = 1}^{k_j}\varphi_2[i/x] \big) \land \varphi_3[k_j/x] \]
According to Definition 3.6:
\[ s.pc \triangleq (\bigvee_{j = 1}^{n} k = k_j) \land \varphi_1 \land (\forall i. \ 1 \leq i \leq k \rightarrow \varphi_2[i/x]) \land \varphi_3[k/x] \]

First, we prove that $m \models s'.pc$ iff $m[k \mapsto \tilde{k}] \models s.pc$ for some $\tilde{k} \in \Nat$. \\
\bm{$\Rightarrow:$}\\
If $m \models s'.pc$, then there exists $1 \leq j \leq n$ such that:
\[ m \models \treepc(n_j) \]
Let $m' \triangleq m[k \mapsto k_j]$.
We will show now that $m' \models s.pc$.
Clearly, $m' \models k = k_j$, so:
\[ m' \models \bigvee_{j = 1}^{n} k = k_j \]
Note that $\treepc(n_j)$ can be rewritten using quantifiers:
\[ \treepc(n_j) \equiv \varphi_1 \land (\forall i. \ (1 \leq i \leq k_j \rightarrow \varphi_2[i/x])) \land \varphi_3[k_j/x] \]
where $i$ is a fresh variable.
As $m \models \treepc(n_j)$ and $m' \models k = k_j$, and $k$ does not appear in $\varphi_1, \varphi_2(x), \varphi_3(x)$:
\[ m' \models \varphi_1 \land (\forall i. \ (1 \leq i \leq k \rightarrow \varphi_2[i/x])) \land \varphi_3[k/x] \]
and consequently:
\[ m' \models (\bigvee_{j = 1}^{n} k = k_j) \land \varphi_1 \land (\forall i. \ (1 \leq i \leq k \rightarrow \varphi_2[i/x])) \land \varphi_3[k/x] \]
\bm{$\Leftarrow:$}\\
If there exists $\tilde{k} \in \Nat$ such that $m[k \mapsto \tilde{k}] \models s.pc$, then there exists $1 \leq j \leq n$ such that:
\[ m[k \mapsto \tilde{k}](k) = m[k \mapsto \tilde{k}](k_j) \]
so:
\[ m[k \mapsto \tilde{k}] \models \varphi_1 \land (\forall i. \ (1 \leq i \leq k_j \rightarrow \varphi_2[i/x])) \land \varphi_3[k_j/x] \]
and in particular (as $k$ does not appear in the formula above):
\[ m \models \varphi_1 \land (\forall i. \ (1 \leq i \leq k_j \rightarrow \varphi_2[i/x])) \land \varphi_3[k_j/x] \]
As mentioned before:
\[ \treepc(n_j) \equiv \varphi_1 \land (\forall i. \ (1 \leq i \leq k_j \rightarrow \varphi_2[i/x])) \land \varphi_3[k_j/x] \]
so $m \models \treepc(n_j)$ and therefore:
\[ m \models \bigvee_{j = 1}^{n} \treepc(n_j) \]

Second, we prove that if $m \models s.pc$ then for every variable $v$ in the symbolic store it holds that:
\[ m(s.\mathit{mem}(v)) = m(s'.\mathit{mem}(v)) \]

Suppose that $m \models s.pc$, and let $v$ be a variable in the symbolic store.
The interesting case is when the merged value of $v$ is encoded without \ite's.
That is, when there exists a term $t(x)$ with a free variable $x$ such that:
\[ t[k_j/x] \eqsyn n_j.s.mem(v) \ \ \ \text{(for every $j = 1,\dots,n$)} \]
and the value of $v$ in $s$ is encoded as:
\[ s.\mathit{mem}(v) \triangleq t[k/x] \]
We already proved that if $m \models s.pc$ then there must exist $1 \leq j \leq n$ such that:
\[ m \models \treepc(n_j) \text{ and } m(k) = m(k_j) \]
Recall that $s'.\mathit{mem}(v)$ is defined by:
\[ ite(\treepc(n_1), n_1.mem(v), ite(\treepc(n_2), n_2.mem(v), \ldots)) \]
which can be rewritten as:
\[ ite(\treepc(n_1), t[k_1/x], ite(\treepc(n_2), t[k_2/x], \ldots)) \]
Recall that $\{\treepc(n_j)\}_{j = 1}^{n}$ correspond to path conditions in the execution tree $t$, which are pairwise unsatisfiable, so:
\[ m(ite(\treepc(n_1), t[k_1/x], ite(\treepc(n_2), t[k_2/x], \ldots))) = m(t[k_j/x]) \]
and since $m(k) = m(k_j)$, we get:
\[ m(t[k_j/x]) = m(t[k/x]) \]
$\blacksquare$

\subsection{Proof of Lemma 3.8}
\subsubsection*{Proof of Lemma \textbf{3.8.1}}
The proof will be done by induction on the length of the hash, which is a sequence of numbers.
In the base case, the length of the hash is 1, that is:
\[ |h(\treepath(n_1))| = |h(\treepath(n_2))| = 1 \]
Both paths $\treepath(n_1)$ and $\treepath(n_2)$ start from the root $r$, so it must hold that:
\[ n_1 = n_2 = r \]
In the induction step, we assume that:
\[ h(\treepath(n_1)) = h(\treepath(n_2)) = h_1 \ldots h_{n - 1}h_n \text{ where $h_i \in \Nat$ } \]
Let $n'_1$ and $n'_2$ be the nodes preceding $n_1$ and $n_2$, respectively, that is:
\[ \treepath(n_1) = \treepath(n'_1); [n_1], \ \treepath(n_2) = \treepath(n'_2); [n_2] \]
where $;$ denotes the concatenation of sequences.
We know that:
\[ h(\treepath(n'_1)) = h(\treepath(n'_2)) = h_1 \ldots h_{n - 1} \]
so by the induction hypothesis:
\[ n'_1 = n'_2 \]
Thus, we can conclude that $n_1$ and $n_2$ have the same parent node,
\ie $n_1$ and $n_2$ are sibling nodes.
Note that:
\[ h(\treepath(n_1)) = h(\treepath(n'_1))h(n_1), \ \ h(\treepath(n_2)) = h(\treepath(n'_2))h(n_2) \]
so it must hold that:
\[ h(n_1) = h(n_2) \]
We assumed that $h$ is valid in $t$ (Definition 3.1),
so if $n_1 \neq n_2$, then $h(n_1) \neq h(n_2)$.
Therefore, $n_1 = n_2$.

\subsubsection*{Proof of Lemma \textbf{3.8.2}}
If $h(\treepath(n_1))$ is a prefix of $h(\treepath(n_2))$, then there exists $\omega \in \Nat^{*}$ such that:
\[ h(\treepath(n_2)) = h(\treepath(n_1)); \omega \]
According to the definition of $h$,
there exists a node $n$ on the path $\treepath(n_2)$ such that:
\[ h(\treepath(n)) = h(\treepath(n_1)) \]
According to 3.8.1, this means that $n$ must be $n_1$, so there is a path $\treepath(n_1, n_2)$.
Each edge in the execution tree $t$ goes from the parent node to the child node,
so the path $\treepath(n_1, n_2)$ is unique.

\subsection{Proof of Lemma 3.10}
First, we prove the following lemma: \\
\textit{Lemma:}
If $h(\treepc(n)) = \omega_1\ldots\omega_n$, then:
\begin{align*}
\treepctail(n) \eqsyn & \ extract(\omega_1) \ \land \\
                      & \ extract(\omega_1, \omega_1\omega_2) \ \land \\
                      & \ \ldots \ \land \\
                      & \ extract(\omega_1\ldots\omega_{n - 1}, \omega_1\ldots\omega_n)
\end{align*}

\textit{Proof of Lemma}:
The proof is done by induction on $n$.
In the base case, $h(\treepath(n)) = \omega_1$, so by the definition of \textit{extract}:
\[ \treepctail(n) \eqsyn extract(\omega_1) \]
In the induction step, we assume that:
\[ h(\treepath(n)) = \omega_1\ldots\omega_{n - 1}\omega_n \]
By the definition of $h$, there exists a node $n'$ such that:
\[ h(\treepath(n')) = \omega_1\ldots\omega_{n - 1} \]
According to the induction hypothesis:
\begin{align*}
\treepctail(n') \eqsyn & \ extract(\omega_1) \ \land \\
                       & \ extract(\omega_1, \omega_1\omega_2) \ \land \\
                       & \ ... \ \land \\
                       & \ extract(\omega_1\ldots\omega_{n - 2}, \omega_1\ldots\omega_{n - 1})
\end{align*}
and according to the definition of \textit{extract}:
\[ \treepctail(n', n) \eqsyn extract(\omega_1\ldots\omega_{n - 1}, \omega_1\ldots\omega_n) \]
Finally, from the definition of $\treepctail$:
\begin{align*}
\treepctail(n) \eqsyn & \ \treepctail(n') \land \treepctail(n', n) \\
               \eqsyn & \ extract(\omega_1) \ \land \\
                      & \ extract(\omega_1, \omega_1\omega_2) \ \land \\
                      & \ ... \ \land \\
                      & \ extract(\omega_1\ldots\omega_{n - 2}, \omega_1\ldots\omega_{n - 1}) \\
                      & \ extract(\omega_1\ldots\omega_{n - 1}, \omega_1\ldots\omega_n)
\end{align*}

%
Now, we go back to the proof of Lemma 3.10.
We assumed that $\{(n_j, k_j)\}_{j = 1}^{n}$ match $(\omega_1, \omega_2, \omega_3)$:
\[ h(\treepath(n_j)) \triangleq \omega_1\omega_2^{k_j}\omega_3 \ \ \ (j = 1, ..., n) \]
By the previous lemma:
\begin{align*}
\treepc(n_j) \eqsyn \ & extract(\omega_1) \ \land \\
                      & ... \\
                      & extract(\omega_1\omega_2^{k_j - 1}, \omega_1\omega_2^{k_j}) \ \land \\
                      & extract(\omega_1\omega_2^{k_j}, \omega_1\omega_2^{k_j}\omega_3)
\end{align*}
and we also assumed that:
\[
\begin{array}{l}
\varphi_1 \eqsyn extract(\omega_1) \\
\varphi_2[i/x] \eqsyn extract(\omega_1\omega_2^{i - 1}, \omega_1\omega_2^{i}) \ \ \ (i = 1, ..., \mathit{max}\{k_j\}_{j = 1}^{n}) \\
\varphi_3[k_j/x] \eqsyn extract(\omega_1\omega_2^{k_j}, \omega_1\omega_2^{k_j}\omega_3) \ \ \ (j = 1, ..., n)
\end{array}
\]
so:
\[ \treepc(n_j) \eqsyn \varphi_1 \land \bigwedge_{i = 1}^{k_j} \varphi_2[i/x] \land \varphi_3[k_j/x] \ \ \ (j = 1, ..., n) \]
and therefore $\{(\treepc(n_j), k_j)\}_{j = 1}^{n}$ match $(\varphi_1, \varphi_2(x), \varphi_3(x))$.
\\
$\blacksquare$


\section{Pseudocode For The Solving Procedure}\label{sec:app-solving}

\textit{Notations.}
We assume closed formulas $\varphi = \bigwedge \clause$ 
where each clause $\clause$ is either a quantifier free formula $\qfform$ or
a universal formula of the form $\qclause$ where  $\bodyform$ is a quantifier free formula with free variable~$\freevar$.
We denote by $\quantified{\varphi}$ resp. $\qfree{\varphi}$ the set of quantified resp. quantifier-free clauses of $\varphi$.
We refer to a pair $(a, e)$ comprised of an array $a$ and an index term $e$ as an \emph{access pair}.
We denote by $\reads(\clause) \triangleq \{ (a, e) \mid \aaccess{a}{e} \in \clause \}$
the set of all access pairs coming from array access terms in $\clause$
and by $\qreads(\clause) \triangleq \{ (a, e) \in \reads(\clause) \mid \freevar \in \fv{e} \}$
the access pairs of a quantified clause $\clause$ in which the index term contains a quantified variable.
We denote by $\arrays(\varphi) \triangleq \bigcup_{\clause \in \quantified{\varphi}} \{ a \mid (a, e) \in \qreads(\clause) \}$
the arrays accessed using a quantified variable.
Given a model $m$ and an access pair $(a, e)$, we define $m(a,e) \triangleq (m(a), m(e))$,
and refer to it as a \emph{semantic access pair}.
We extend this notation to sets of such pairs in a point-wise manner.

\textit{Solving procedure.}
Our solving procedure is given in \Cref{alg:solve}.
Its main function is \computemodel{} which works in four stages. 

\textit{(1) Quantifier stripping by formula weakening (\cref{line:solve:strip,line:solve:smt,line:solve:strip-succ}).}
\computemodel{} starts by invoking  $\strip(\varphi)$
which weakens $\varphi$ into a quantifier free formula $\varphi_{\mathit{QF}}$ by replacing quantified clauses with implied quantifier free clauses.
Specifically, each quantified clause $\qclause$ in $\varphi$ is replaced with quantifier free clauses stating that
(a) the instantiation of $\bodyform$ to $\boundvar=1$, denoted $\bodyform[1/\boundvar]$, must hold if $0<k$,
and (b) if $\neg\bodyform[t/i]$ holds for some term $t$ then $t$ cannot be in the range $[1,k]$.
If the SMT solver fails to find a model for $\varphi_{\mathit{QF}}$ than  $\varphi$ is also unsatisfiable.
If a model was found, we check, optimistically, whether it is also a model of $\varphi$.

\begin{example}\label{ex:app-strip}
\revision{
Consider the following query, a simplification of a representative query from our experiments:
\begin{equation*}
\begin{split}
\varphi \triangleq & (s[n] = 0) \land (1 \leq k \leq 10) \land (s[k - 1] = 8) \land \\
                   & (\forall i. \ 1 \leq i \leq k \rightarrow s[i - 1] \neq 0)
\end{split}
\end{equation*}
Note that (a) the instantiation of the quantified formula using $i = 1$ results in $k \geq 1 \rightarrow s[0] \neq 0$,
and (b) $s[n] = 0$ is obtained by substituting $\neg(s[i - 1] \neq 0)[n + 1/i]$.
Thus, the weakened query obtained by quantifier stripping is given by:
\begin{equation*}
\begin{split}
\varphi_{\mathit{QF}} \triangleq & (s[n] = 0) \land (1 \leq k \leq 10) \land (s[k - 1] = 8) \land \\
                                 & (k \geq 1 \rightarrow s[0] \neq 0) \land \neg(1 \leq n + 1 \leq k)
\end{split}
\end{equation*}
The following model, for example, is a model of $\varphi_{\mathit{QF}}$:
\[ m \triangleq \{n \mapsto 7, k \mapsto 7, s \mapsto [1,0,0,0,0,0,8,0]\} \]
but, unfortunately, is not a model of $\varphi$.
}

\revision{
Note that if we would consider a different model of $\varphi_{\mathit{QF}}$:
\[ m \triangleq \{n \mapsto 1, k \mapsto 1, s \mapsto [8,0]\} \]
then we could get a satisfying model of $\varphi$.
}

\end{example}

\textit{(2) Assignment Duplication (\cref{line:solve:duplicate,line:solve:duplicate-succ}).\footnote{
We explain the role of \textit{conflicts} in the next stage, for now, assume that it is an empty set.
}}
If $m$ is not a model of $\varphi$, we use \duplicatemodel{} to modify $m$ into in a model $m_d$ which 
assigns to every array cell accessed by a quantified clause a value of a cell in that array that was explicitly constrained  by $\varphi_{\mathit{QF}}$.
To do so, {\duplicatemodel} iterates over all the quantified clauses $\qclause$ of $\varphi$,
and for every array $a$ that $\bodyform$ accesses using the quantified variable $i$ it
\begin{inparaenum}[(i)]
\item records in $r$ the set of access pairs coming from such accesses (\cref{alg:duplicate:r}),
\item nondeterministically chooses one of these pairs $(a, e)$ (\cref{alg:duplicate:chosen}), and
\item determines the value $\semanticterm{v}$ stored in $a$ at the chosen index $e$ when $\boundvar$ is substituted by $1$.
\end{inparaenum}
Recall that accesses to $e[1/\boundvar]$ were explicitly constrained by $\varphi_{\mathit{QF}}$
due to the added instantiations in \cref{alg:strip:inst-one}.
Hence, the value $\semanticterm{v}$ assigned to them by $m$ is a good candidate to fill in all the other array cells of $a$ constrained by $\varphi$.
Accordingly, the interpretation of $\select$ in $m$ is modified
such that every semantic access pair pertaining to the access pair $(a, e)$ is mapped to $\semanticterm{v}$ (\cref{alg:duplicate:update}).
The duplication, however, is rather naive and might result in a model which does not even satisfy $\varphi_{\mathit{QF}}$.

\begin{example}\label{ex:app-duplicate}
\revision{
Continuing \Cref{ex:app-strip},
we pick from the quantified clause the accessed offset $i - 1$ of the array $s$,
and update the value of $s[j]$ to $m(s[i - 1][1/i])$ for each $1 \leq j \leq 6$.
This results in the following model:
\[ m_d \triangleq \{ n \mapsto 7, k \mapsto 7, s \mapsto [1,1,1,1,1,1,1,0] \} \]
The model $m_d$ helps to satisfy the quantified clause,
but does not satisfy $\varphi$ due to the violation of the clause $s[k - 1] = 8$.
}

\revision{
Note that if we would consider a different model of $\varphi_{\mathit{QF}}$ in the stripping stage:
\[ m \triangleq \{n \mapsto 7, k \mapsto 7, s \mapsto [8,0,0,0,0,0,8,0]\} \]
then the model $m_d$ obtained after assignment duplication could be:
\[ m_d \triangleq \{ n \mapsto 7, k \mapsto 7, s \mapsto [8,8,8,8,8,8,8,0] \} \]
which does satisfy $\varphi$.
}
\end{example}


\begin{algorithm}[t]
\caption{A specialized solving procedure}
\label{alg:solve}\label{alg:strip}\label{alg:assignment-duplication}\label{alg:model-repair}
\begin{algorithmic}[1]

\Function{\strip}{$\varphi$}
    \State $\varphi_s \gets \textit{true}$
    \For {$(\qclause) \in \quantified{\varphi}$}
        \State $\varphi_s \gets \varphi_s \land  (k \geq 1 \rightarrow \bodyform[1/\freevar]) \land {} \label{alg:strip:inst-one}$ \\
               \hspace{6.8ex} $\big(\bigwedge \{ \neg (1 \leq t \leq k) \mid (\neg \bodyform[t/\freevar]) \in \qfree{\varphi} \}\big)$
    \EndFor
    \State \Return $\big(\bigwedge \qfree{\varphi}\big) \land \varphi_s$
\EndFunction

\Function{\duplicatemodel}{$\varphi, m, \conflicts$}
    \For {$(\qclause) \in \quantified{\varphi}$} \label{alg:duplicate:iter}
        \For {$a \in  \arrays(\bodyform)$}
            \State $r \gets \{ (a', e) \in \qreads(\bodyform) \mid a' = a \}$ \label{alg:duplicate:r}
            \State $\textit{let } (a, e) \in r$ \label{alg:duplicate:chosen}
            \State $\semanticterm{v} \gets m[\freevar \mapsto 1](\aaccess{a}{e})$
            \For {$2 \leq n \leq m(k)$}
                \State $\semanticterm{a}, \semanticterm{o} \gets m[\freevar \mapsto \mathit{n}](a, e)$
                \If {$(\semanticterm{a}, \semanticterm{o}) \not\in \mathit{conflicts}$}
                    \State $m \gets \updateselect(m, \semanticterm{a}, \semanticterm{o}, \semanticterm{v})$ \label{alg:duplicate:update}
                \EndIf
            \EndFor
        \EndFor
    \EndFor
    \State \Return $m$
\EndFunction

\Function{\repairmodel}{$\varphi, m$}
    \State $\mathit{conflicts} \gets \emptyset$, $\map \gets []$
    \For {$(\qclause) \in \quantified{\varphi}$} \label{alg:repair:loop1-start}
        \For {$1 \leq \mathit{n} \leq m(k)$}
            \If {$m[\freevar \mapsto \mathit{n}] \not\models \bodyform$}
                \State $\mathit{conflicts} \gets \mathit{conflicts} \cup m[\freevar \mapsto \mathit{n}](\qreads(\bodyform))$  \label{alg:repair:unsatqclause}
            \EndIf
        \EndFor
    \EndFor \label{alg:repair:loop1-end}

    \For {$\qfclause \in \qfree{\varphi}$} \label{alg:repair:loop2-start}
        \If {$m \not\models \qfclause$}
            \State $\mathit{conflicts} \gets \mathit{conflicts} \cup m(\text{reads}(\qfclause))$     \label{alg:repair:unsatqfclause}
        \EndIf
    \EndFor \label{alg:repair:loop2-end}

    \For {$(\qclause) \in \quantified{\varphi}$} \label{alg:repair:loop3-start}
        \For {$1 \leq \mathit{n} \leq m(k)$}
            \dtx{why intersect with conflicts?}
            \For {$(\semanticterm{a},\semanticterm{o}) \in m[\freevar \mapsto \mathit{n}](\qreads(\bodyform)) \cap \mathit{conflicts} $}
                \State $map[(\semanticterm{a}, \semanticterm{o})] \gets map[(\semanticterm{a}, \semanticterm{o})] \cup \{\bodyform[\mathit{n}/\freevar]\}$
            \EndFor
        \EndFor
    \EndFor \label{alg:repair:loop3-end}

    \State $\varphi' \gets \strip(\varphi) \land
            \big(\bigwedge_{(\semanticterm{a}, \semanticterm{o}) \in \mathit{conflicts}} map[(\semanticterm{a}, \semanticterm{o})]\big)$
    \label{alg:repair:strip}
    \State $r \gets \{ \aaccess{a}{e} \in \termsform(\varphi) \mid a \not \in \arrays(\varphi) \}$ \label{alg:repair:terms}
    \State $m' \gets \smtcomputemodel(\varphi' \land \big(\bigwedge_{t \in r} t = m(t)\big))$ \label{alg:repair:compute}

    \If {$m' = \bot $} \Return $\bot$
    \EndIf
    \State \Return $\duplicatemodel(\varphi, m', \mathit{conflicts})$ \label{alg:repair:duplicate}

\EndFunction

\Function{compute-model}{$\varphi$}
    \State $\varphi_{\mathit{QF}} \gets \strip(\varphi)$\label{line:solve:strip}
    \State $m \gets \smtcomputemodel(\varphi_{\mathit{QF}})$\label{line:solve:smt}
    \If {$m= \bot \lor m \models \varphi$} \Return $m$ \label{line:solve:strip-succ}
    \EndIf
    \State $m_{d} \gets \duplicatemodel(\varphi, m,\emptyset)$\label{line:solve:duplicate}
    \If {$m_{d} \models \varphi$} \Return $m_{d}$\label{line:solve:duplicate-succ}
    \EndIf
    \State $m_{r} \gets \repairmodel(\varphi, m_{d})$ \label{line:solve:repair}
    \If {$m_{r} \neq \bot \land m_{r} \models \varphi$}  \Return $m_{r}$ \label{line:solve:repair-succ}
    \EndIf
    \State \Return $\smtcomputemodel(\varphi)$ \label{line:solve:fallback}
\EndFunction

\end{algorithmic}
\end{algorithm}

\textit{(3) Model Repair (\cref{line:solve:repair,line:solve:repair-succ}).}
If $m_d$ is  not a model of $\varphi$, we invoke \repairmodel{}  to further modify $m_d$ into another model, 
$m_r$, which, much like $m_d$, attempts to satisfy the constraints on the contents of arrays that are imposed by $\varphi$ but omitted in $\varphi_{\mathit{QF}}$.
However, it does so in a more principled way than \duplicatemodel{}:
Firstly, \repairmodel{} collects a set of semantic access pairs, called $\mathit{conflicts}$,
from clauses that are not satisfied by $m_d$ (lines \ref{alg:repair:loop1-start}-\ref{alg:repair:loop2-end}).
This set is used both to identify quantifier-free constraints that need to be added to $\varphi_{\mathit{QF}}$, and to later avoid overwriting ``good'' array contents. 
Secondly, \repairmodel{} iterates again over the quantified clauses $\qclause$ of $\varphi$
and collects for each semantic access pair $(\semanticterm{a}, \semanticterm{o})$ in $\mathit{conflicts}$
the set of all instantiations that constrain it (lines \ref{alg:repair:loop3-start}-\ref{alg:repair:loop3-end}).
These instantiations are implied by $\varphi$ in all models that agree with $m_d$ on the value of $k$, which are our focus.
Thirdly, $\varphi_{\mathit{QF}}$ is strengthened into $\varphi'$ by conjoining it with the collected instantiations (\cref{alg:repair:strip}).
Fourthly, we obtain a modification of $m_d$ that satisfies $\varphi'$ (rather than computing a new model from scratch)
by further strengthening $\varphi'$ with constraints that force the interpretation of closed terms to agree with their interpretation in $m_d$
(\cref{alg:repair:compute}).
The exception is terms of the form $\aaccess{a}{e}$ where $a \in \arrays(\varphi)$ (\cref{alg:repair:terms}),
for which a new interpretation is sought.
Finally, if a model $m'$ is found, then duplication is applied to obtain $m_r$.
However, this time the semantic access pairs in $\mathit{conflicts}$, which were explicitly constrained when computing $m'$,
are excluded from the duplication in order to avoid their overwriting.

\textit{(4) Fallback (\cref{line:solve:fallback}).}
If no model $m_r$ is found, or if it does not satisfy $\varphi$,
we ask the SMT solver to find a model for $\varphi$.


\section{Evaluation}

\subsection{Additional Experimental Results}

\subsubsection{Metrics}
Recall that when we reported in the body of the paper each of the metrics (analysis time and coverage),
we focused on the relevant subsets of subjects (APIs and whole programs):
When reporting speedups,
we did not consider subjects in which both modes timed out, as there was no meaningful speedup to report,
and when reporting coverage,
we did not consider subjects in which both modes terminated, as they trivially obtained the same coverage.
For completeness, we report the obtained results when all the subjects are considered.
More specifically,
\Cref{table:app-vs-cfg} corresponds to~\Cref{table:vs-cfg},
\Cref{table:app-scale} corresponds to~\Cref{table:scale},
\Cref{table:app-vs-base} corresponds to~\Cref{table:vs-base},
\Cref{table:app-solving-procedure} corresponds to~\Cref{table:solving-procedure},
and \Cref{table:app-incremental} corresponds to~\Cref{table:incremental}.
Qualitatively, the results are similar to the ones discussed in the body of the paper,
however, the obtained medians often got cluttered to 1x (speedup) and 0\% (coverage) due to the presence of many trivial comparisons.

\begin{table}[t]
\caption{Comparison of \patternabv vs. \cfg.}
\label{table:app-vs-cfg}
\centering
{\footnotesize
\begin{tabular}{|l|r|r|r|r|}
\cline{1-5}
\multicolumn{1}{|c|}{} &
\multicolumn{2}{c|}{\text{Speedup ($\times$)}} &
\multicolumn{2}{c|}{\text{Coverage (\%)}} \\
\multicolumn{1}{|c|}{} &
\multicolumn{1}{c|}{Avg.} & \multicolumn{1}{c|}{Med.} &
\multicolumn{1}{c|}{Avg.} & \multicolumn{1}{c|}{Med.} \\

\hline
\libosip & 2.46 & 1.00 & 16.36 & 0.00 \\
\hline
\wget & 1.42 & 1.00 & 11.63 & 0.00 \\
\hline
\libtasn & 0.97 & 0.96 & 0.00 & 0.00 \\
\hline
\libpng & 0.97 & 1.00 & 1.86 & 0.00 \\
\hline
\revisionrow \apr & 1.87 & 1.00 & 0.83 & 0.00 \\
\hline
\revisionrow \jsonc & 2.51 & 2.59 & 0.33 & 0.00 \\
\hline
\revisionrow \busybox & 1.32 & 1.00 & -0.82 & 0.00 \\
\hline
\end{tabular}
}
\end{table}

\begin{table}[t]
\caption{Comparison of \patternabv vs. \cfg under different capacity settings (column \textit{Capacity}) in \libosip.}
\label{table:app-scale}
\centering
{\footnotesize
\begin{tabular}{|l|r|r|r|r|}
\cline{1-5}
\multicolumn{1}{|c|}{\text{Capacity}} &
\multicolumn{2}{c|}{\text{Speedup ($\times$)}} &
\multicolumn{2}{c|}{\text{Coverage (\%)}} \\
\multicolumn{1}{|c|}{} &
\multicolumn{1}{c|}{Avg.} & \multicolumn{1}{c|}{Med.} &
\multicolumn{1}{c|}{Avg.} & \multicolumn{1}{c|}{Med.} \\

\hline
10 & 2.46 & 1.00 & 16.36 & 0.00 \\
\hline
20 & 1.76 & 1.00 & 19.40 & 9.09 \\
\hline
50 & 1.26 & 1.00 & 13.02 & 2.43 \\
\hline
100 & 1.16 & 1.00 & 8.77 & 1.61 \\
\hline
200 & 1.25 & 1.00 & 3.62 & 0.00 \\
\hline
\end{tabular}
}
\end{table}

\begin{table}[t]
\caption{Comparison of \patternabv vs. \base.}
\label{table:app-vs-base}
\centering
{\footnotesize
\begin{tabular}{|l|r|r|r|r|}
\cline{1-5}
\multicolumn{1}{|c|}{} &
\multicolumn{2}{c|}{\text{Speedup ($\times$)}} &
\multicolumn{2}{c|}{\text{Coverage (\%)}} \\
\multicolumn{1}{|c|}{} &
\multicolumn{1}{c|}{Avg.} & \multicolumn{1}{c|}{Med.} &
\multicolumn{1}{c|}{Avg.} & \multicolumn{1}{c|}{Med.} \\

\hline
\libosip & 3.23 & 1.00 & 9.14 & 0.00 \\
\hline
\wget & 1.48 & 1.00 & -1.79 & 0.00 \\
\hline
\libtasn & 2.36 & 1.51 & 0.80 & 0.00 \\
\hline
\libpng & 1.08 & 1.00 & 21.63 & 6.63 \\
\hline
\revisionrow \apr & 2.89 & 1.00 & -0.05 & 0.00 \\
\hline
\revisionrow \jsonc & 1.28 & 1.00 & 0.33 & 0.00 \\
\hline
\revisionrow \busybox & 0.87 & 1.00 & -1.00 & 0.00 \\
\hline
\end{tabular}
}
\end{table}

\begin{table}[t]
\caption{Impact of solving procedure.}
\label{table:app-solving-procedure}
\centering
{\footnotesize
\begin{tabular}{|l|r|r|r|r|r|r|}
\cline{1-7}
\multicolumn{1}{|c|}{} &
\multicolumn{2}{c|}{\text{Speedup ($\times$)}} &
\multicolumn{4}{c|}{\text{Coverage (\%)}} \\

\multicolumn{1}{|c|}{} &
\multicolumn{2}{c|}{} &
\multicolumn{2}{c|}{\text{Line}} &
\multicolumn{2}{c|}{\text{Path}} \\

\multicolumn{1}{|c|}{} &
\multicolumn{1}{c|}{Avg.} & \multicolumn{1}{c|}{Med.} &
\multicolumn{1}{c|}{Avg.} & \multicolumn{1}{c|}{Med.} &
\multicolumn{1}{c|}{Avg.} & \multicolumn{1}{c|}{Med.} \\

\hline
\libosip & 1.22 & 1.00 & 0.14 & 0.00 & 48.48 & 0.43 \\
\hline
\wget & 1.67 & 1.00 & 12.90 & 0.00 & 95.95 & 10.05 \\
\hline
\libtasn & 0.99 & 0.99 & 0.00 & 0.00 & -0.34 & 0.00 \\
\hline
\libpng & 1.00 & 1.00 & -0.21 & 0.00 & 2.41 & 0.00 \\
\hline
\revisionrow \apr & 1.69 & 1.00 & 0.08 & 0.00 & 19.15 & 0.00 \\
\hline
\revisionrow \jsonc & 2.34 & 2.16 & 0.00 & 0.00 & 15.98 & 0.00 \\
\hline
\revisionrow \busybox & 1.23 & 1.00 & 0.40 & 0.00 & 7.31 & 0.00 \\
\hline
\end{tabular}
}
\end{table}

\begin{table}[t]
\caption{Impact of incremental state merging.}
\label{table:app-incremental}
\centering
{\footnotesize
\begin{tabular}{|l|r|r|r|r|}
\cline{1-5}
\multicolumn{1}{|c|}{} &
\multicolumn{2}{c|}{\text{Speedup ($\times$)}} &
\multicolumn{2}{c|}{\text{Coverage (\%)}} \\
\multicolumn{1}{|c|}{} &
\multicolumn{1}{c|}{Avg.} & \multicolumn{1}{c|}{Med.} &
\multicolumn{1}{c|}{Avg.} & \multicolumn{1}{c|}{Med.} \\

\hline
\libosip & 2.39 & 1.00 & 15.18 & 0.00 \\
\hline
\wget & 0.99 & 1.00 & 10.75 & 0.00 \\
\hline
\libtasn & 0.98 & 1.00 & 0.00 & 0.00 \\
\hline
\libpng & 0.99 & 1.00 & 2.15 & 0.00 \\
\hline
\revisionrow \apr & 1.29 & 1.00 & 0.95 & 0.00 \\
\hline
\revisionrow \jsonc & 1.01 & 1.00 & 0.00 & 0.00 \\
\hline
\revisionrow \busybox & 0.96 & 1.00 & 0.51 & 0.00 \\
\hline
\end{tabular}
}
\end{table}

\subsubsection{\patternabv vs. \base}
\begin{figure*}
    \centering
    \begin{subfigure}[b]{0.30\textwidth}
        \centering
        \includegraphics[width=\textwidth]{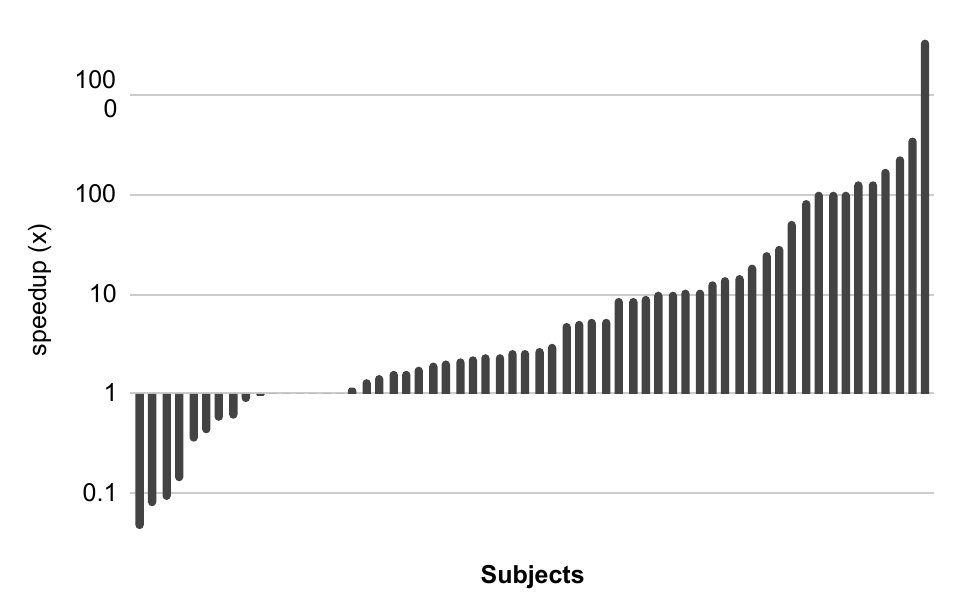}
        \caption{Speedup in analysis time ($\times$) in the subjects where at least one of the modes terminated (in log-scale).}
        \label{fig:base-speedup}
    \end{subfigure}
    \hfill
    \begin{subfigure}[b]{0.30\textwidth}
        \centering
        \includegraphics[width=\textwidth]{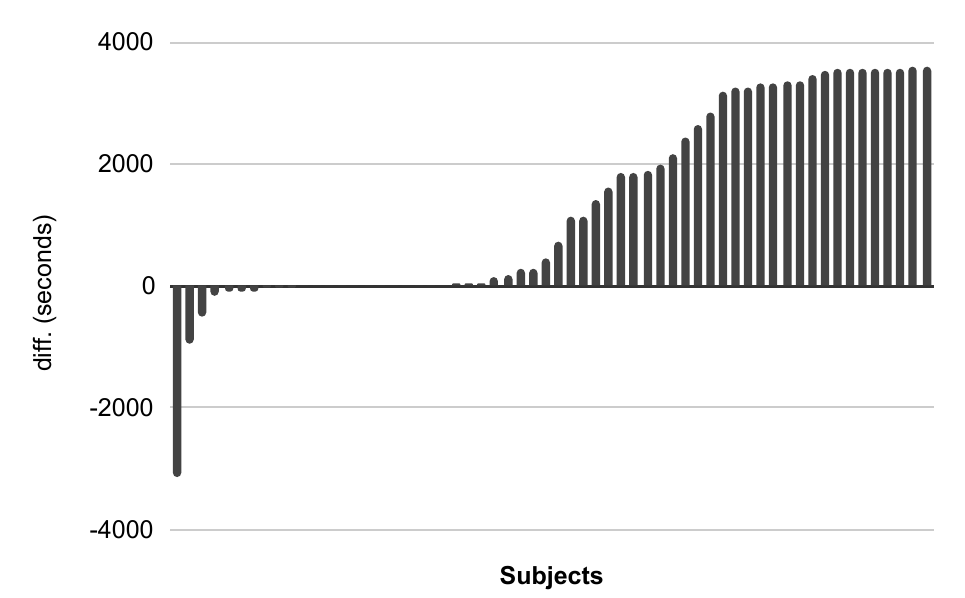}
        \caption{Difference in analysis time (\emph{seconds}) in the subjects where at least one of the modes terminated.}
        \label{fig:base-diff}
    \end{subfigure}
    \hfill
    \begin{subfigure}[b]{0.30\textwidth}
        \centering
        \includegraphics[width=\textwidth]{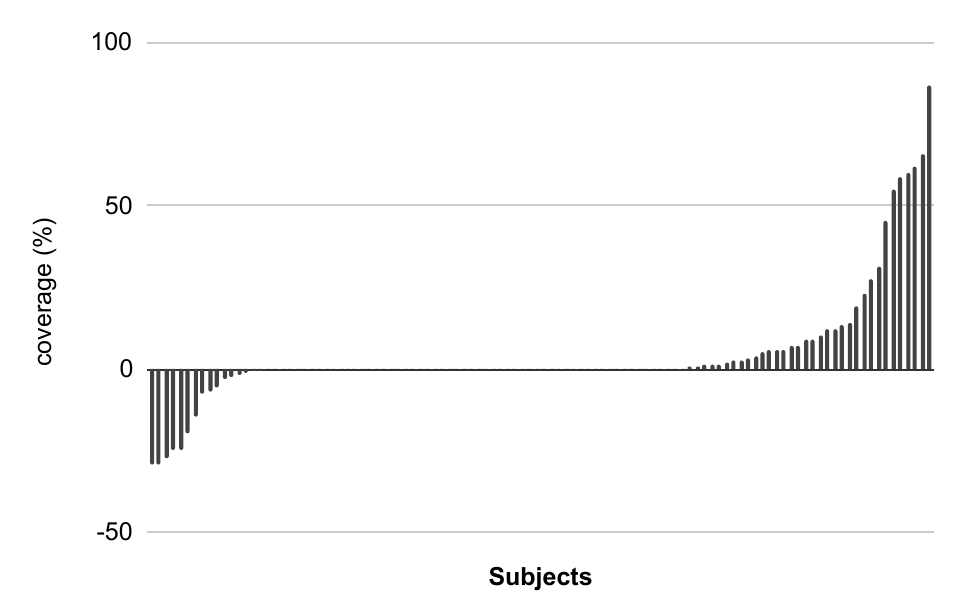}
        \caption{Relative increase in coverage (\%) in the subjects where at least one of the modes timed out.}
        \label{fig:base-coverage}
    \end{subfigure}
    \caption{Breakdown of the improvement of \patternabv over \base per subject.}
    \label{fig:base-breakdown}
\end{figure*}

The breakdown of improvement of \patternabv over \base per subject is shown in \Cref{fig:base-breakdown}.

\subsection{Solving Procedure}

In this section,
we perform some additional experiments to evaluate the different aspects of our solving procedure.

First, we evaluate the effectiveness of our solving procedure by checking its success rate.
The results are shown in \Cref{table:effectiveness-solving-procedure}.
Column \textit{Total} shows the total number of generated quantified queries,
and column \textit{Solved} shows the percentage of queries that were solved by our solving procedure.
The results show that the solving procedure was able to handle most of the generated queries.
In addition, we measured the individual contribution of the different stages of our solving procedure:
quantifier stripping, assignment duplication, and model repair.
The results, shown in the appendix (Section C.2) for space reasons,
indicate that each stage plays a part in the overall efficacy of the procedure.

In addition, we show the contribution of the different stages in our solving procedure:
quantifier stripping (\optstrip), assignment duplication (\optdup), and model repair (\optrepair).
\Cref{table:hits} shows the number of solved quantified queries in each of the stages:
quantifier stripping only (\optstrip),
quantifier stripping and assignment duplication (\optstrip + \optdup),
and the complete algorithm (\optstrip + \optdup + \optrepair).
Column \textit{Fallback} shows the number of quantified queries that our solving procedure failed to solve.

\begin{table}[t]
\caption{Effectiveness of solving procedure.}
\label{table:effectiveness-solving-procedure}
\centering
{\footnotesize
\begin{tabular}{|l|r|r|r|r|r|r|r|r|}
\cline{1-9}
\multicolumn{1}{|c|}{} &
\multicolumn{1}{c|}{\text{Total}} &
\multicolumn{1}{c|}{\text{Solved (\%)}} \\

\hline
\revisionrow \libosip & 517026 & 94 \\
\hline
\revisionrow \wget & 208535 & 98 \\
\hline
\revisionrow \libtasn & 44 & 100 \\
\hline
\revisionrow \libpng & 2411 & 86 \\
\hline
\revisionrow \apr & 187700 & 99 \\
\hline
\revisionrow \jsonc & 7390 & 98 \\
\hline
\revisionrow \busybox & 58013 & 98 \\
\hline
\end{tabular}
}
\end{table}

\begin{table}[t]
\caption{The number of solved queries in the different stages of the solving procedure.}
\label{table:hits}
\centering
{\footnotesize
\begin{tabular}{|l|r|r|r|r|}
\cline{1-5}
\multicolumn{1}{|c|}{} &
\multicolumn{1}{c|}{\text{\optstrip}} &
\multicolumn{1}{c|}{\text{\optstrip + \optdup}} &
\multicolumn{1}{c|}{\text{\optstrip + \optdup + \optrepair}} &
\multicolumn{1}{c|}{\textit{Fallback}} \\
\hline
\libosip & 453688 & 1361 & 33868 & 28109 \\
\hline
\wget & 187175 & 3759 & 15241 & 2360 \\
\hline
\libtasn & 44 & 0 & 0 & 0 \\
\hline
\libpng & 2077 & 2 & 0 & 332 \\
\hline
\revisionrow \apr & 135664 & 48014 & 3829 & 193 \\
\hline
\revisionrow \jsonc & 6205 & 246 & 859 & 80 \\
\hline
\revisionrow \busybox & 51917 & 4837 & 260 & 999 \\
\hline
\end{tabular}
}
\end{table}

\revision{
\subsection{Concrete-Size Model}
We performed an additional experiment using the concrete-size memory model.
The results are shown in \Cref{table:app-vs-cfg-no-symbolic-size}.
As can be seen,
\patternabv achieved better results than \cfg in most of the benchmarks.
}

\revision{
As for the comparison between the results in this experiment
and the results obtained with the symbolic-size memory model (\Cref{table:vs-cfg} in the paper):
In terms of analysis time,
here we had more speedup in \libosip and less speedup in other benchmarks.
In terms of coverage,
here we had less improvement in \libosip, more improvement in \wget,
and the results in other benchmarks were similar.
}

\revision{
\subsection{Default Search Heuristic}
We performed an additional experiment using \klee's default search heuristic.\footnote{
When state merging is enabled, the default search heuristic is set using the command-line option: \texttt{-search=nurs:covnew}.
}
The results are shown in \Cref{table:app-vs-cfg-coverage} and \Cref{table:app-vs-base-coverage}.
}

\revision{
In general,
when the analysis achieves full exploration with one search heuristic,
the analysis time with other search heuristics is usually similar.
Indeed, as can be seen from the results,
the analysis times here are very similar to those obtained using the DFS search heuristic (\Cref{table:vs-cfg} and \Cref{table:vs-base}  in the paper).
}


\revision{
In terms of coverage,
the results here are comparable to those obtained using the DFS search heuristic.
In some cases (for example, \busybox and \libpng) we had more improvement,
and in other cases (for example, \libosip and \wget) we had less improvement.
}

%
%
%

\begin{table*}[t]
\caption{Comparison of \patternabv vs. \cfg without the symbolic-size model.}
\label{table:app-vs-cfg-no-symbolic-size}
\centering
{\footnotesize
\begin{tabular}{|l|r|r|r|r|r|r|}
\cline{1-7}
\multicolumn{1}{|c|}{} &
\multicolumn{3}{c|}{\text{Speedup ($\times$)}} &
\multicolumn{3}{c|}{\text{Coverage (\%)}} \\
\multicolumn{1}{|c|}{} &
\multicolumn{1}{c|}{\#} & \multicolumn{1}{c|}{Avg.} & \multicolumn{1}{c|}{Med.} &
\multicolumn{1}{c|}{\#} & \multicolumn{1}{c|}{Avg.} & \multicolumn{1}{c|}{Med.} \\

\hline
\revisionrow \libosip & 28/35 & 10.08 & 8.13 & 20/35 & 16.82 & 2.32 \\
\hline
\revisionrow \wget & 14/31 & 2.01 & 1.43 & 21/31 & 25.78 & 0.00 \\
\hline
\revisionrow \libtasn & 7/13 & 0.96 & 0.96 & 6/13 & 0.00 & 0.00 \\
\hline
\revisionrow \libpng & 3/12 & 0.96 & 0.98 & 9/12 & -0.55 & 0.00 \\
\hline
\revisionrow \apr & 10/20 & 1.97 & 1.23 & 11/20 & 2.67 & 0.00 \\
\hline
\revisionrow \jsonc & 4/5 & 0.95 & 1.50 & 1/5 & 0.00 & 0.00 \\
\hline
\revisionrow \busybox & 10/30 & 1.29 & 1.00 & 23/30 & -1.33 & 0.00 \\
\hline
\end{tabular}
}
\end{table*}

\begin{table*}[t]
\caption{Comparison of \patternabv vs. \cfg with the \textit{nurs:covnew} search heuristic.}
\label{table:app-vs-cfg-coverage}
\centering
{\footnotesize
\begin{tabular}{|l|r|r|r|r|r|r|}
\cline{1-7}
\multicolumn{1}{|c|}{} &
\multicolumn{3}{c|}{\text{Speedup ($\times$)}} &
\multicolumn{3}{c|}{\text{Coverage (\%)}} \\
\multicolumn{1}{|c|}{} &
\multicolumn{1}{c|}{\#} & \multicolumn{1}{c|}{Avg.} & \multicolumn{1}{c|}{Med.} &
\multicolumn{1}{c|}{\#} & \multicolumn{1}{c|}{Avg.} & \multicolumn{1}{c|}{Med.} \\

\hline
\revisionrow \libosip & 17/35 & 6.69 & 4.00 & 28/35 & 15.12 & 2.11 \\
\hline
\revisionrow \wget & 10/31 & 2.85 & 1.91 & 25/31 & 11.66 & 0.00 \\
\hline
\revisionrow \libtasn & 7/13 & 0.94 & 0.96 & 6/13 & -0.42 & 0.00 \\
\hline
\revisionrow \libpng & 1/12 & 0.74 & 0.74 & 11/12 & 9.39 & 13.04 \\
\hline
\revisionrow \apr & 10/20 & 3.60 & 1.72 & 11/20 & 1.55 & 0.00 \\
\hline
\revisionrow \jsonc & 4/5 & 3.10 & 2.84 & 1/5 & 0.82 & 0.82 \\
\hline
\revisionrow \busybox & 8/30 & 1.69 & 1.14 & 23/30 & 3.03 & 0.00 \\
\hline
\end{tabular}
}
\end{table*}

\begin{table*}[t]
\caption{Comparison of \patternabv vs. \base with the \textit{nurs:covnew} search heuristic.}
\label{table:app-vs-base-coverage}
\centering
{\footnotesize
\begin{tabular}{|l|r|r|r|r|r|r|}
\cline{1-7}
\multicolumn{1}{|c|}{} &
\multicolumn{3}{c|}{\text{Speedup ($\times$)}} &
\multicolumn{3}{c|}{\text{Coverage (\%)}} \\
\multicolumn{1}{|c|}{} &
\multicolumn{1}{c|}{\#} & \multicolumn{1}{c|}{Avg.} & \multicolumn{1}{c|}{Med.} &
\multicolumn{1}{c|}{\#} & \multicolumn{1}{c|}{Avg.} & \multicolumn{1}{c|}{Med.} \\

\hline
\revisionrow \libosip & 18/35 & 9.58 & 3.02 & 27/35 & 8.52 & 0.00 \\
\hline
\revisionrow \wget & 10/31 & 2.87 & 5.29 & 26/31 & -5.30 & 0.00 \\
\hline
\revisionrow \libtasn & 7/13 & 4.80 & 9.29 & 7/13 & 0.65 & 0.00 \\
\hline
\revisionrow \libpng & 1/12 & 4.80 & 4.80 & 11/12 & 2.11 & 3.50 \\
\hline
\revisionrow \apr & 10/20 & 8.12 & 4.51 & 15/20 & -0.14 & 0.00 \\
\hline
\revisionrow \jsonc & 4/5 & 1.36 & 3.59 & 2/5 & 0.41 & 0.41 \\
\hline
\revisionrow \busybox & 10/30 & 2.11 & 2.01 & 23/30 & 5.57 & 0.00 \\
\hline
\end{tabular}
}
\end{table*}

\end{document}